\documentclass[12pt,preprint]{aastex}
\slugcomment{To be submitted to The Astrophysical Journal}

\begin{document}
\shortauthors{Sepinsky, Willems, \& Kalogera}
\shorttitle{Equipotential surfaces and Lagrangian points in eccentric binaries}
\title{Equipotential Surfaces and Lagrangian points in Non-synchronous, 
Eccentric Binary and Planetary Systems}
\author{J. F. Sepinsky, B. Willems, V. Kalogera}
\email{j-sepinsky, b-willems, and vicky@northwestern.edu}
\affil{Department of Physics and Astronomy, Northwestern University,
2145 Sheridan Road, Evanston, IL 60208}


\begin{abstract}

We investigate the existence and properties of equipotential surfaces
and Lagrangian points in non-synchronous, eccentric binary star and
planetary systems under the assumption of quasi-static equilibrium. We
adopt a binary potential that accounts for non-synchronous rotation and
eccentric orbits, and calculate the positions of the Lagrangian points
as functions of the mass ratio, the degree of asynchronism, the orbital
eccentricity, and the position of the stars or planets in their relative
orbit. We find that the geometry of the equipotential surfaces may
facilitate non-conservative mass transfer in non-synchronous, eccentric
binary star and planetary systems, especially if the component stars or
planets are rotating super-synchronously at the periastron of their
relative orbit.  We also calculate the volume-equivalent radius of the
Roche lobe as a function of the four parameters mentioned above.
Contrary to common practice, we find that replacing the radius of a
circular orbit in the fitting formula of \citet{1983ApJ...268..368E}
with the instantaneous distance between the components of eccentric
binary or planetary systems does not always lead to a good approximation
to the volume-equivalent radius of the Roche-lobe. We therefore provide
generalized analytic fitting formulae for the volume-equivalent
Roche lobe radius appropriate for non-synchronous, eccentric binary star
and planetary systems.  These formulae are accurate to better than 1\% 
throughout the relevant 2-dimensional parameter space that covers a 
dynamic range of 16 and 6 orders of magnitude in the two dimensions.

\end{abstract}



\section{Introduction}
\label{sec-introduction}

The Roche model has served for a long time as a fundamental tool in the
study of the interactions and observational characteristics of the
components of gravitational two-body systems. Among the most noteworthy
applications are the modeling of the shapes of binary components, the
study of mass transfer in interacting binaries, and the orbital
stability of satellites in planetary systems. At the basis of these
applications are the properties of the potential governing the
gravitational and centrifugal forces operating in the system. In
particular, the shape of equipotential surfaces and the existence of
stationary points (where the net force exerted on a test particle
vanishes) play a central role in our understanding of the evolution of
binary star and planetary systems. 

More often than not, applications of the Roche model are built on the
assumption that the orbit of the system is circular and that the system
components are rotating synchronously with the orbital motion. Under
this assumption, the components can be treated as static with respect to
a co-rotating frame of reference and their shapes are determined by
equipotential surfaces of the system. However, the assumption of
circular orbits and synchronous rotation cannot always be justified,
neither on observational nor on theoretical grounds. 

Observational support for non-circular orbits and non-synchronous
rotation is very well established at present. Catalogs of eclipsing
binaries as early as those by \cite{1913ApJ....38..158S} list binaries
with non-negligible eccentricities ranging from 0.01 to 0.14. The sample
as well as the largest measured orbital eccentricity has since increased
drastically \citep[e.g.][]{1950QB801.S8......., 1999AJ....117..587P,
1997ApJS..113..367B, 2005ASPC..333...88H, 2004A&A...424..727P} and now
also includes binaries in the Small and Large Magellanic Clouds
\citep[e.g.][]{2002MNRAS.331..609B, 2003A&A...402..509C,
2004MNRAS.349..833B, 2005MNRAS.357..304H}. Evidence for non-synchronous
rotation of the components of close binary systems is equally abundant,
and points to both sub- and supersynchronously rotating component stars
\citep[e.g.][]{1950QB801.S8......., 1974A&A....35..259L,
1989A&A...211...56H, 2006astro.ph..8154M, 1990AJ....100.1981V}.

Non-circular orbits are also observed in extrasolar planetary systems.
Since the discovery of the first extrasolar planet orbiting a solar-type
star, 51\,Peg\,, \citep{1995Natur.378..355M, MB}, the sample of known
exoplanets has grown to more than 200, many of which orbit their host
star with non-negligible orbital eccentricities.  The most eccentric
exoplanetary orbit known to date is that of HD\,20782b which has an
eccentricity of 0.9 \citep{2006MNRAS.369..249J}.  Constraints on
exoplanetary rotation rates are yet to be inferred from observations.

Theoretically, circularization and synchronization processes are driven
by tidal interactions between the components of binary or planetary
systems\footnote{For very close systems, orbital angular momentum losses
due to gravitational wave emission may also contribute to the
circularization process \citep{1964PhRv..136.1224P}.}. Predicted
circularization and synchronization time scales, however, depend
strongly on the initial system parameters and theoretical uncertainties
in the strength of tidal dissipation mechanisms
\citep{2004ApJ...602L.121M, 2005ApJ...620..970M, 2006astro.ph..8154M}.
In addition, since synchronization tends to occur faster than
circularization, components of binary and planetary systems tend to
become synchronized with the orbital motion near periastron before the
completion of circularization. The stellar and planetary rotation rates
then deviate from synchronism at all other orbital phases.

Such deviations from synchronism in gravitational two-body systems
induce time-dependent oscillations in the atmospheres of the component
stars or planets \citep{1970A&A.....4..452Z, 1975A&A....41..329Z}.
Therefore these objects can no longer be treated as static with respect
to a rotating frame of reference, unless the time scale of the
oscillatory motions is sufficiently long compared to the dynamical time
scale of the star or planet. The validity of this approximation was
first discussed by \citet{1963ApJ...138.1112L} for the case of
non-synchronous circular binaries \citep[see
also][]{1963AcA....13..106K, 1978A&A....62..317S}. He concluded that as
long as the rotational angular velocities of the stars do not deviate
considerably from synchronicity, the components may be approximated as
static and their shapes may be determined by the instantaneous
equipotential surfaces of the binary.  \citet{1963ApJ...138.1112L} also
found the approximation to be valid in the limiting case of stars
rotating with angular velocities close to the break-up angular velocity
due to the dominance of the centrifugal force over the gravitational
forces. 

\citet{Plavec58} was the first to abandon the assumption of synchronous
rotation and study the effects of the asynchronicity on the shapes of
the binary components and the position of the inner Lagrangian point
$L_1$ (the stationary point of the Roche potential located in between
the two stars on the line connecting their centers of mass) in binaries
with circular orbits. He found that the position of the $L_1$ point
tends to move closer to the mass center of a rapidly rotating binary
component with increasing values of its rotational angular velocity.
Correspondingly, the critical radius at which one component starts to
transfer mass to its companion decreases with increasing rotational
angular velocity. Later investigations extended the work of
\citet{Plavec58} by more carefully considering the validity of the
assumptions underlying the Roche model \citep{1963ApJ...138.1112L,
1963AcA....13..106K} and considering the effect of spin-orbit
misalignment in circular binaries \citep{1982ApJ...257..703A}.
\citet{1979ApJ...229.1008L} also incorporated the effects of gas
dynamics and heat transport in the surface layers of the binary
components.

\cite{1976ApJ...209..574A} further generalized the Roche model to
account for eccentric binary orbits with the aim of modeling the light
and radial-velocity variations observed in the eclipsing binary pulsar
Vela X-1. His generalized binary potential was subsequently used by
\citet{1979ApJ...234.1054W} to discuss the computation of light and
radial-velocity curves of eccentric binaries with non-synchronously
rotating component stars. \citet{2005MNRAS.358..544R}, performed
smoothed particle hydrodynamics calculations to validate the application
of the Roche model to eccentric binaries with component stars rotating
close to synchronism with the orbital angular velocity at periastron. 
Part of their work, however, is flawed due to an incorrect term in the
expression for the gravitational potential (see \S~\ref{sec-massloss}
for details). 

Despite the vast observational support for the occurrence of eccentric
orbits in binary and exoplanetary systems, a detailed account of the
properties of the generalized Roche potential used to study these
systems seems to be lacking from the literature.  The geometry of the
equipotential surfaces as well as the existence, location, and potential
height of Lagrangian points are necessary, for instance, in determining
when mass transfer is initiated in non-synchronous, eccentric binaries
along with the determining how mass and angular momentum flow during
these mass-transfer phases. Consequently the theoretical study of
component interactions in non-synchronous binaries of arbitrary
eccentricity requires the understanding of this geometry and properties
of the binary potential. Such interactions (e.g., mass exchange, mass
and angular momentum loss) very often occur as binaries evolve and their
modeling enters almost all population synthesis calculations of binaries
of current interest (X-ray binaries, gamma-ray burst progenitors, binary
compact objects, etc.). They are also relevant to the evolution of
planetary systems along with the possibility of mass transfer and loss
from planets in orbit around their host stars. The pursuit of
theoretical understanding of these interaction processes in the context
of broad binary evolution studies is what motivates the study presented
here. 

The plan of the paper is as follows. In \S\ref{sec-potential}, we
introduce the Roche potential describing the motion of mass elements in
non-synchronous, eccentric binary and planetary systems. In
\S\ref{sec-validity}, we discuss the time-dependence of the potential
and investigate the conditions under which the potential varies sufficiently
slowly to be considered quasi-static. Adopting these conditions, the
existence and stability of stationary (Lagrangian) points of the
potential are examined in \S\ref{sec-lpoints0}. In \S\ref{sec-veqrlr},
we calculate the volume-equivalent radius of the Roche Lobe as a
function of the system parameters and provide a generalization of the
\citet{1983ApJ...268..368E} fitting formula appropriate for
non-synchronous, eccentric binary star and planetary systems. In
\S\ref{sec-massloss}, we examine the height of the potential along the
line connecting the mass centers of the stellar or planetary binary
components and discuss the implications for mass transfer and mass loss
from the system. The final section, \S\ref{sec-conclusion}, is devoted
to concluding remarks.


\section{The Roche Potential for Non-Synchronous Eccentric Binaries}
\label{sec-potential}

We consider a binary system of stars\footnote{For brevity, we will
simply refer to the system components as stars, bearing in mind that the
analysis is valid for planetary systems as well.} with masses $M_1$ and
$M_2$ orbiting one another in a fixed Keplerian orbit with period
$P_{\rm orb}$, semi-major axis $a$, and eccentricity $e$.  The distance
$D$ between the mass centers of the two stars and the Keplerian orbital
angular velocity $\omega_K$ of the stars are, at any time $t$, given by
\begin{equation}
\label{eq-D}
D(t) = \frac{a\left(1-e^2\right)}{1+e\cos{\nu}},
\end{equation}
and
\begin{equation}
\omega_K(t) = \frac{2\pi}{P_{\rm orb}} 
\frac{\left( 1+e\cos{\nu} \right)^2}{\left(1-e^2\right)^{3/2}},
\end{equation}
where $\nu$ is the true anomaly.

We assume the first star (hereafter star~1) rotates
uniformly\footnote{The assumption of uniform rotation is made because it
is the simplest possible, though it is not without merit. 
\citet{1998Natur.393..139S} and \citet{1999A&A...341L...1S}
argue that magnetic fields could establish uniform rotation.  On the 
other
hand, there are compelling cases of single and binary stars for which
differential rotation has been claimed \citep[See,
e.g.,][]{1976ApJ...205..462P, 2003A&A...407L..47D}.}  about an axis
perpendicular to the orbital plane with a fixed rotational angular
velocity $\vec{\Omega}_1$ parallel to and directed in the same sense as
the orbital angular velocity, $\vec{\omega}_K$. We furthermore assume
star~1 is sufficiently centrally condensed to be approximated by a point
mass surrounded by a uniformly rotating zero-density envelope.  We
assume this envelope is convectively stable so that no bulk motions of
the mass elements of star~1 occur other than those due to the star's
rotation and the tidal force exerted by the companion. The companion
star (hereafter star~2) is treated as a point mass.

\clearpage
\begin{figure}
\plotone{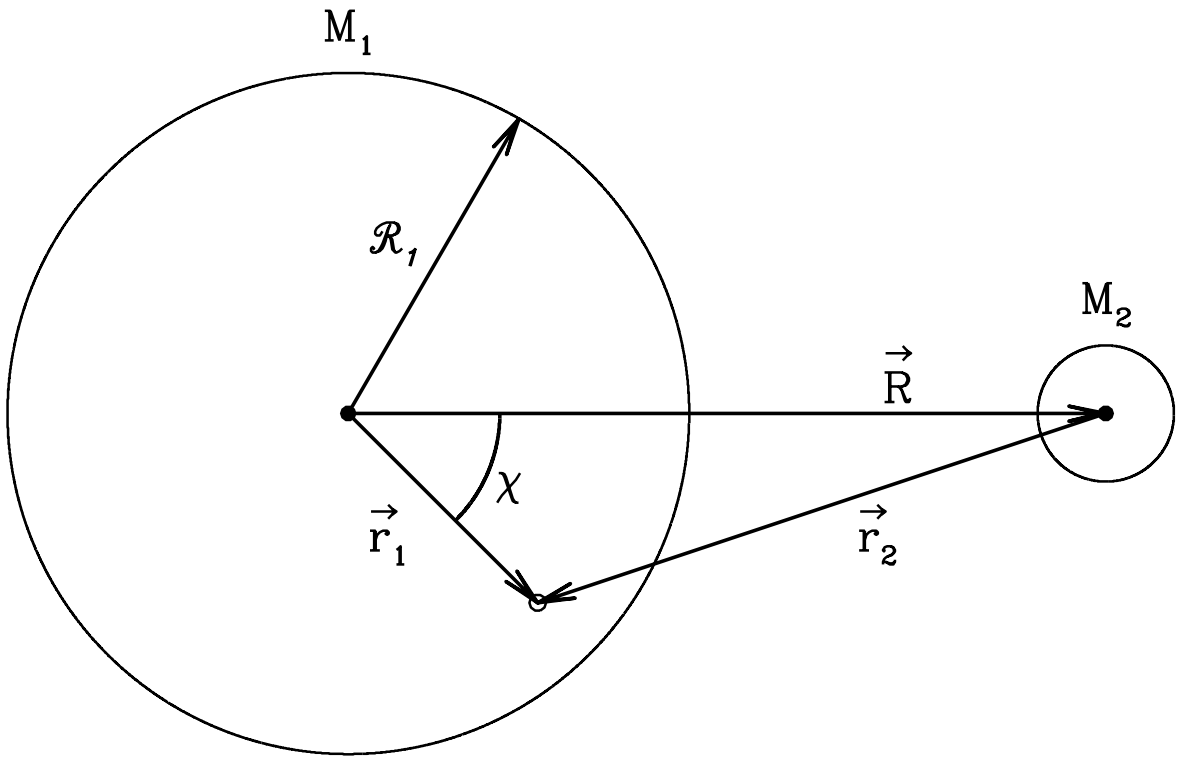}
\caption{Schematic representation of the position vectors used in the
derivation of the equation of motion of a mass element (open circle) in
a component of a non-synchronous eccentric binary.}
\label{fig-coords}
\end{figure}
\clearpage
The mass elements of star~1 are, at any given time, subjected to a sum
of gravitational and rotational forces caused by the stars' mutual
gravitational attraction, rotation, and orbital motion.  We describe the
motion of the mass elements of star~1 with respect to a Cartesian
coordinate frame $OXYZ$ with origin $O$ located at the center of mass of
star~1 and $Z$-axis parallel to $\vec{\Omega}_1$.  We furthermore let
the frame rotate about the $Z$-axis with the rotational angular velocity
of star~1.  The equation of motion of a mass element of star~1 with
respect to $OXYZ$ is then given by
\begin{equation}
\label{eq-r1ddot}
\ddot{\vec{r}}_1 = -\frac{1}{\rho}\vec{\nabla} P -
\vec{\nabla} V_1-2\, \vec{\Omega}_1\times\dot{\vec{r}}_1,
\end{equation}
where $\vec{r}_1$ is the position vector of the mass element with
respect to the mass center of star~1, $\rho$ and $P$ are the mass
density and pressure of the mass element, and
\begin{equation}
\label{eq-V1}
V_1 = -G\frac{M_1}{|\vec{r}_1|} - G\frac{M_2}{|\vec{r}_2|} -
\frac{1}{2}|\vec{\Omega}_1|^2\left(X^2 + Y^2\right) + G\frac{M_2X}{D^2}
\end{equation}
is the potential giving rise to the gravitational and centrifugal forces
acting on the mass element \citep[for details, see][and references
therein]{SWK2006}. Here, $G$ is the Newtonian constant
of gravitation, $\vec{r}_2$ is the position vector of the considered
mass element with respect to the mass center of star~2, and $X$ and $Y$ are
the Cartesian coordinates of the mass element in the plane of the orbit.  

We note that Eq.~(\ref{eq-V1}) is valid only when the $X$-axis of the
reference frame coincides with the line connecting the mass centers of
the binary components. Since the initial orientation of the $X$- and
$Y$-axes in the orbital plane is arbitrary, the $X$-axis may be chosen
to coincide with the line connecting the mass centers of the component
stars at any given time without loss of generality.  The definitions of
the position vectors introduced in Eqs.~(\ref{eq-r1ddot})
and~(\ref{eq-V1}) are shown schematically in Fig.~\ref{fig-coords}. 

The potential $V_1$ can be rewritten by expanding the gravitational
attraction exerted by star~2 on the considered mass element of star~1 in
terms of Legendre polynomials $P_n(x)$ as
\begin{equation}
\label{eq-ex}
-G\frac{M_2}{|\vec{r}_2|} = -G\frac{M_2}{|\vec{R}|}
\sum_{n=0}^\infty\left(\frac{|\vec{r}_1|}{|\vec{R}|}\right)^n
P_n\left( \cos{\chi} \right)
\end{equation}
where $\vec{R}=\vec{r}_1-\vec{r}_2$ is the position vector of the mass
center of star~2 with respect to the mass center of star~1, and $\chi$
is the angle between $\vec{R}$ and $\vec{r}_1$. The first term in the
expansion of $-G M_2/ |\vec{r}_2|$ is independent of the position of the
considered mass element and thus does not contribute to the force
exerted on the element. The second term in the expansion of $-G M_2/
|\vec{r}_2|$ is identical to the last term in Eq.~(\ref{eq-V1}).  
Equation~(\ref{eq-r1ddot}) is therefore equivalent to
\begin{equation}
\label{eq-r1ddot2}
\ddot{\vec{r}}_1 = -\frac{1}{\rho}\vec{\nabla} P -
\vec{\nabla} V_1^\ast-2\, \vec{\Omega}_1\times\dot{\vec{r}}_1,
\end{equation}
where
\begin{equation}
\label{eq-V1epsT}
V_1^\ast = -G\frac{M_1}{|\vec{r}_1|} - \frac{1}{2}|\vec{\Omega}_1|^2(X^2 +
Y^2) + W(\vec{r}_1,t).
\end{equation}
In the latter equation, $W(\vec{r}_1,t)$ is the tide-generating
potential defined as
\begin{equation}
\label{eq-epsT}
W(\vec{r}_1,t) = -G\frac{M_2}{|\vec{R}|}
\sum_{n=2}^\infty\left(\frac{|\vec{r}_1|}{|\vec{R}|}\right)^n
P_n\left( \cos{\chi} \right)
\end{equation}
\citep[e.g.,][]{1970A&A.....4..452Z, 1990A&A...237..110P}.

Denoting the radius of star~1 by ${\cal R}_1$ and restricting the
tide-generating potential to its dominant $n=2$ terms, it follows from
dimensional analysis that
\begin{equation}
\frac{{(1/2) |\vec{\Omega}_1|^2 \left( X^2 + Y^2 \right)}}{ 
  {G M_1/|\vec{r}_1|}} \sim \left( \frac{{|\vec{\Omega}_1|}}{{\Omega_c}} 
  \right)^2,
\end{equation}
and
\begin{equation}
\label{eq-o4}
\frac{W}{{G M_1/|\vec{r}_1|}} \sim \left( \frac{{\cal R}_1}{D} \right)^3
  \frac{M_2}{M_1},
\end{equation}
where $\Omega_c \sim (G M_1/{\cal R}_1^3)^{1/2}$ is the break-up angular
velocity of star~1. The centrifugal distortion thus becomes
increasingly important when $|\vec{\Omega}_1| \rightarrow \Omega_c$,
while the tidal distortion becomes important when ${\cal R}_1
\rightarrow D$ and $M_2 \rightarrow M_1$.


\section{The quasi-static approximation}
\label{sec-validity}

For a binary with a circular orbit in which star~1 is rotating
synchronously with the orbital motion, the potential $V_1$ is
independent of time and the mass elements of star~1 are at rest with
respect to the co-rotating frame of reference. 
Equation~(\ref{eq-r1ddot}) therefore reduces to the condition for
hydrostatic equilibrium
\begin{equation}
\label{eq-PV}
\frac{1}{\rho}\vec{\nabla} P = -\vec{\nabla} V_1,
\end{equation}
so that surfaces of constant pressure and density coincide with surfaces
of constant $V_1$.  The possible shapes of star~1 are thus determined by 
the equipotential surfaces of the binary.

For a binary with an eccentric orbit or non-synchronously rotating
component stars, $V_1$ is a periodic function of time. At any given
time, the mass elements of star~1 therefore undergo tidally induced
oscillations on a time scale $\tau_{\rm tide}$ determined by the
difference between the instantaneous orbital angular velocity
$\omega_K(t)$ and the rotational angular velocity $|\vec{\Omega}_1|$:
\begin{equation}
\tau_{\rm tide} = {{2\pi} \over{|\omega_K(t) - |\vec{\Omega}_1||}}.
\end{equation}
From dimensional analysis, it follows that
\begin{equation}
\label{eq-rddotdim}
\frac{| \ddot{\vec{r}}_1 |}{| (\vec{\nabla} P)/\rho|} \sim \frac{| 
\ddot{\vec{r}}_1 |}{| \vec{\nabla}V_1 |} \sim 
  \frac{\tau_{\rm dyn}^2}{\tau_{\rm tide}^2},
\end{equation}
\begin{equation}
\label{eq-omegardotdim}
\frac{| \vec{\Omega}_1 \times \dot{\vec{r}}_1 |}{| (\vec{\nabla} 
P)/\rho|} \sim 
\frac{| \vec{\Omega}_1 \times \dot{\vec{r}}_1 |}{| 
\vec{\nabla}V_1 |} \sim  \frac{|\vec{\Omega}_1|}{\Omega_c} \,
  \frac{\tau_{\rm dyn}}{\tau_{\rm tide}},
\end{equation}
where $\tau_{\rm dyn} = (GM_1/{\cal R}_1^3)^{-1/2}$ is the dynamical
time scale of star~1. Hence, when $\tau_{\rm dyn} \ll \tau_{\rm tide}$,
the $\ddot{\vec{r}}_1$ and $\vec{\Omega}_1 \times \dot{\vec{r}}_1$ terms
in Eq.~(\ref{eq-r1ddot}) are negligible compared to the other terms, so
that Eq.~(\ref{eq-r1ddot}) approximately reduces to Eq.~(\ref{eq-PV}).
In this limiting case, the motions of the mass elements are so slow
that, at each instant, star~1 can be considered to be quasi-static with
respect to the rotating frame of reference $OXYZ$. The instantaneous
shape of the star can then be approximated by the instantaneous surfaces
of constant $V_1$. \citet{1963ApJ...138.1112L} referred to this as the 
{\it first approximation} \citep[see also][]{1978A&A....62..317S, 
1979ApJ...234.1054W}.

The condition that $\tau_{\rm dyn} \ll \tau_{\rm tide}$ can be expressed
in terms of the orbital elements and properties of star~1 by writing
$\tau_{\rm tide}$ as
\begin{equation}
 \label{eq-T} 
 \tau_{\rm tide} = \frac{2\pi}{\omega_P}
 \left|\left(\frac{1+e\cos{\nu}}{1+e}\right)^2-f\right|^{-1},
\end{equation}
where
\begin{equation}
\omega_P=\frac{2\pi}{P_{\rm orb}} 
\frac{\left(1+e\right)^{1/2}}{\left(1-e\right)^{3/2}}
\end{equation}
is the orbital angular velocity at periastron, and
$f=|\vec{\Omega}_1|/\omega_P$ is the rotational angular velocity of star
1 in units of $\omega_P$. It follows that $\tau_{\rm dyn} \ll \tau_{\rm
tide}$ when
\begin{equation}
\label{eq-porb} 
\frac{P_{\rm orb}}{\tau_{\rm dyn}}\gg \alpha(e,f,\nu) ,
\end{equation}
where $\alpha$ is a function of the orbital eccentricity $e$, the ratio
of rotational to orbital angular velocity at periastron $f$, and the
true anomaly $\nu$:
\begin{equation}
\label{eq-porb2} 
\alpha (e,f,\nu) = \frac{\left(1+e\right)^{1/2}}{\left(1-e\right)^{3/2}} 
\left|\left(\frac{1+e\cos{\nu}}{1+e}\right)^2 - f \right|.
\end{equation}
\clearpage
\begin{figure}
\plotone{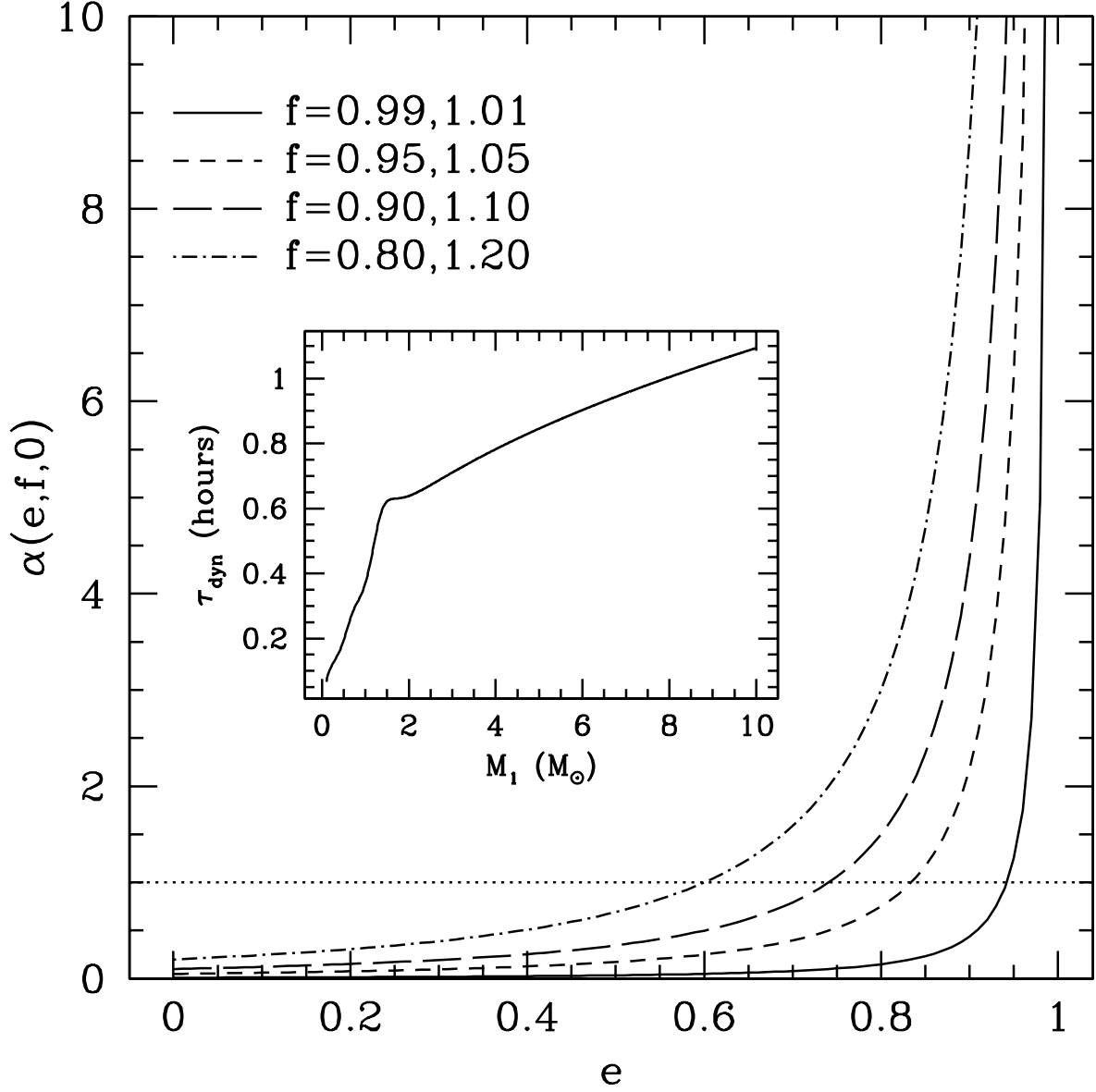}
\caption{Variations of $\alpha(e, f, \nu)$ as a function of $e$, for
$\nu=0$ and $0.8 \le f \le 1.2$. Due to the symmetry of $\alpha(e,f,0)$
about $f=1$, the curves associated with $f=0.80$, 0.90, 0.95, 0.99 are
identical to those associated with $f=1.20$, 1.10, 1.05, 1.01,
respectively.  For $\alpha(e,f,0) \le 1$ (below the dotted line), the 
quasi-static approximation is valid when $P_{\rm orb} \gg \tau_{\rm 
dyn}$.  The inset shows the variation of $\tau_{\rm dyn}$ as a function 
of stellar mass for stars on the ZAMS according to the mass-radius 
relation of \citet{1996MNRAS.281..257T}.}
\label{fig-Pcon}
\end{figure}
\clearpage
When the companion is at the periastron of the relative orbit, $\alpha
(e,f,\nu)$ reduces to
\begin{equation}
\label{eq-porb0} 
\alpha (e,f,0) = \frac{\left(1+e\right)^{1/2}}{\left(1-e\right)^{3/2}} 
\left|1 - f \right|.
\end{equation}
Hence, when $f=1$ and the binary component stars are located at the
periastron of their relative orbit, star~1 can be safely assumed to be
quasi-static for all values of the orbital eccentricity $e < 1$.
However, once $f$ deviates from unity, the rapid increase of $\omega_P$
with increasing orbital eccentricity causes $\alpha(e,f,0)$ to also
increase rapidly with increasing orbital eccentricity.  This is
illustrated in more detail in the main panel of Fig.~\ref{fig-Pcon}. For
the considered range of $f$-values, the function $\alpha(e,f,0)$ is
smaller than 1 for $e \la 0.6$.  Hence, in this parameter range, the
quasi-static approximation is valid at the periastron of the binary
orbit if $P_{\rm orb} \gg \tau_{\rm dyn}$. As is illustrated by the
inset of Fig.~\ref{fig-Pcon}, this condition is easily satisfied for
zero-age main sequence (ZAMS) stars of mass $M_1 \la 10\,M_\odot$,
provided the orbital period is longer than approximately 10\,hr. For
$M_1 \la 1\,M_\odot$ this condition may be relaxed even further.  Due to 
the nonlinear dependence of $\alpha(e,f,\nu)$ on $\nu$, the constraint 
on $P_{\rm orb}$ may be more or less restrictive away from the 
periastron of the relative orbit.


\section{Lagrangian Points}
\label{sec-lpoints0}

\subsection{Existence and Location}
\label{sec-lpoints}

In the case of a binary with a circular orbit and synchronously rotating
component stars, there are five points at which the gravitational forces
exactly balance the centrifugal forces. A mass element with zero
velocity experiences no accelerations at these points and thus remains
stationary with respect to the co-rotating frame of reference $OXYZ$.
Three of these so-called Lagrangian points lie on the line connecting
the mass centers of the binary components, and two lie on the tips of
equilateral triangles whose bases coincide with the line connecting the
mass centers of the component stars.

For binaries with eccentric orbits and/or non-synchronously rotating
component stars, the tide-generating potential introduces a
time-dependence in the potential $V_1$ and thus in the position of the
stationary points (assuming these points still exist). To examine the
existence and properties of these points, we adopt the quasi-static
approximation, $\tau_{\rm dyn} \ll \tau_{\rm tide}$, and introduce the
following dimensionless quantities:
\[ V_D = \frac{V_1}{{G M_2/D}}, \,\,\, q={M_1 \over M_2}, 
  \,\,\, r_D = \frac{{|\vec{r}_1|}}{D}, 
\] 
\[ X_D = \frac{X}{D}, \,\,\, Y_D =
  \frac{Y}{D}, \,\,\, Z_D = \frac{Z}{D}. \]
From Eq.~(\ref{eq-V1}), it follows that
\begin{eqnarray}
\label{eq-VD}
V_D &=& X_D - \frac{q}{r_D} - \frac{1}{\left(r_D^2-2X_D+1\right)^{1/2}} 
\nonumber \\ 
&-& \frac{1}{2} \left(X_D^2 + Y_D^2 \right) \left(1+q\right) {\cal A}(f, 
e, \nu),
\end{eqnarray}
where 
\begin{equation}
{\cal A}(f, e, \nu) = 
\frac{f^2\left(1+e\right)^4}{\left(1+e\cos{\nu}\right)^3}.
\label{eq-A}
\end{equation}
This factor groups all dependencies of the potential on the degree
of asynchronism, the orbital eccentricity, and the mean anomaly. 
In the particular case of a binary with a circular orbit 
and
synchronously rotating component stars, ${\cal A}=1$.

\begin{figure}
\plotone{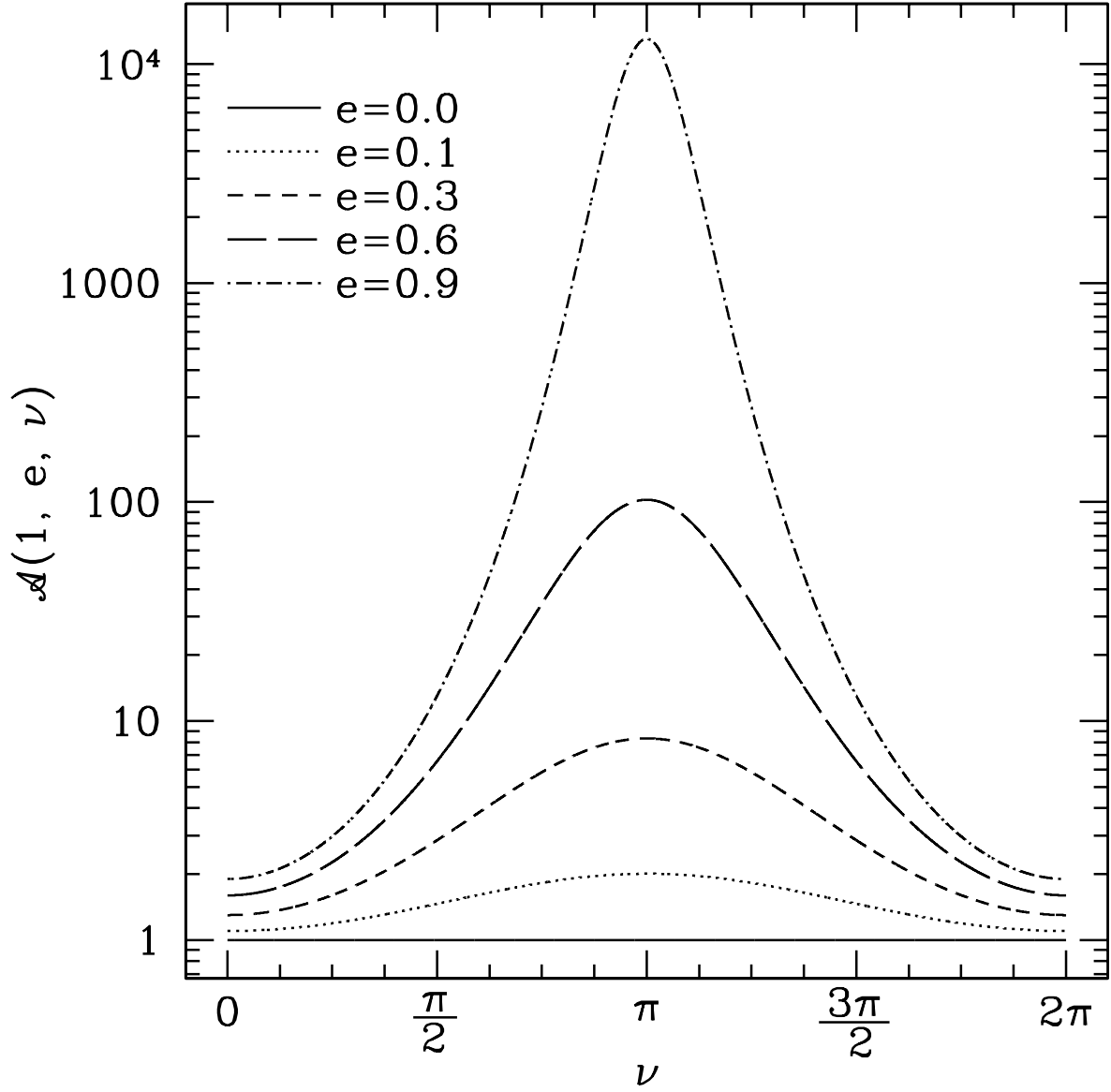}
\caption{Variations of ${\cal A}(f, e, \nu)$ as a function $e$ and
$\nu$, for $f=1$. Variations of ${\cal A}(f, e, \nu)$ for other values
of $f$ are obtained by rescaling the plotted curves by a factor of
$f^2$.}
\label{fig-A}
\end{figure}

The dependence of ${\cal A}(f, e, \nu)$ on $e$ and $\nu$ is illustrated
in Fig.~\ref{fig-A} for $f=1$. Since ${\cal A}(f, e, \nu)$ is linearly
proportional to the square of the rotational angular velocity in units
of the orbital angular velocity at periastron, the curves are easily
rescaled for different values of $f$.  For a given value of $\nu$,
${\cal A}(f, e, \nu)$ increases with increasing $e$, while for a given
$e$, ${\cal A}(f, e, \nu)$ reaches a maximum at apastron and decreases
towards periastron.  ${\cal A}(f, e, \nu)$ is furthermore most sensitive
to the value of the orbital eccentricity near $\nu=\pi$, where it
increases from ${\cal A}(f, e, \nu) = 1$ for $e=0$ to ${\cal A}(f, e,
\nu) \approx 10^4$ for $e=0.9$.  Moreover, at periastron, ${\cal A}(f,
e, \nu) \le 2$ for all $e<1$ and $f \leq 1$.
\clearpage
\clearpage
The stationary points of the binary are solutions of the vector equation
$\vec{\nabla} V_D=0$ with respective $X$, $Y$, and $Z$ components
\begin{eqnarray}
\label{eq-gradvx}
X_D \left[ \frac{q}{r_D^3} \right.&+& 
\frac{1}{\left(r_D^2-2X_D+1\right)^{3/2}} - \left.
(1+q){\cal A}(f, e, \nu) \vphantom{\frac{}{{}^{}_{}}}\right] \nonumber \\ &-& 
\frac{1}{\left(r_D^2-2X_D+1\right)^{3/2}} +1 =  0, 
\end{eqnarray}
\begin{eqnarray}
\label{eq-gradvy}
Y_D \left[ \frac{q}{r_D^3} \right. &+& 
\frac{1}{\left(r_D^2-2X_D+1\right)^{3/2}}
\nonumber \\ &-& \left.
\vphantom{\frac{}{{}^{}_{}}}(1+q){\cal A}(f, e, \nu)\right] = 0, 
\hspace{2cm}
\end{eqnarray}
\begin{equation}
\label{eq-gradvz}
Z_D\left[ \frac{q}{r_D^3} \right. +
\left. \frac{1}{\left(r_D^2-2X_D+1\right)^{3/2}} \right] =0,
  \hspace{1cm}
\end{equation}
where we have used the fact that $r_D^2-2X_D+1 = (X_D-1)^2 + Y_D^2 +
Z_D^2$ is a positive quantity. Since both terms inside the square
brackets in Eq.~(\ref{eq-gradvz}) are positive, the equation can only be
satisfied by setting $Z_D=0$.  Thus, any and all stationary points
necessarily lie in the plane of the orbit, as in the case for circular
binaries with synchronously rotating component stars.  We note in
passing that Lagrangian points outside the orbital plane have been shown
to exist when the spin axis of star~1 is inclined with respect to the
orbital angular momentum axis of the binary \cite[see,
e.g.,][]{1982ApJ...257..703A, 1983ApJ...266..776M}.

Letting $Z_D=0$, we solve Eqs.~(\ref{eq-gradvx}) and~(\ref{eq-gradvy})
for $X_D$ and $Y_D$ by distinguishing between two possible cases: $Y_D =
0$ and $Y_D \ne 0$.

{\em (i) $Y_D = 0$.} Setting $Y_D=Z_D=0$ reduces Eq.~(\ref{eq-gradvx}) 
to
\begin{equation}
\label{eq-L123}
q\,\frac{X_D}{|X_D|^3} + \frac{X_D-1}{\left|X_D-1\right|^3} -
X_D(1+q){\cal A}(f,e,\nu)+1 = 0.
\end{equation}
Leaving aside the trivial solution $X_D=0$, this equation has three real
solutions which can only be determined numerically and which correspond
to three co-linear stationary points on the line connecting the mass
centers of the binary components.  In accordance with the nomenclature
for circular synchronized binaries, we will refer to these points as the
Lagrangian points $L_1$, $L_2$, and $L_3$. The $L_1$ point is the
stationary point located between the two component stars ($0 < X_D <
1$), the $L_2$ point is the stationary point located opposite star~2 as
seen from star~1 ($X_D > 1$), and the $L_3$ point is the stationary
point located opposite star~1 as seen from star~2 ($X_D < 0$).  
\clearpage
\begin{figure}
\plotone{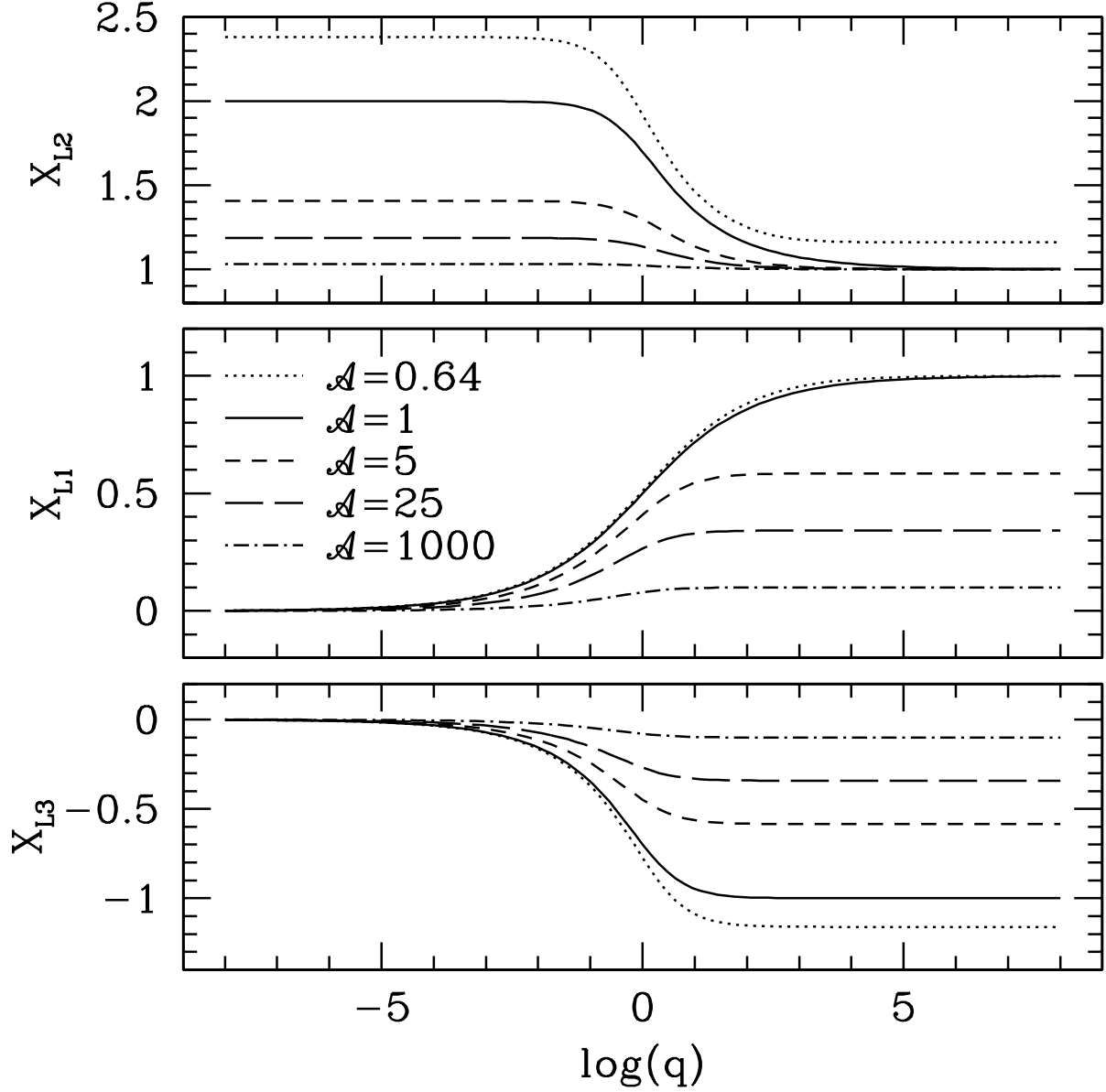}
\caption{The position the co-linear Lagrange points $L_1$, $L_2$,
and $L_3$ along the line connecting the mass centers of the binary
components as a function of the logarithm of the mass ratio, $q$, for
different values of ${\cal A}$.}
\label{fig-lptsA}
\end{figure}
\clearpage
The location of each of the three co-linear Lagrangian points on
the $X$-axis of the $OXYZ$ frame is shown in Fig.~\ref{fig-lptsA} as a
function of $q$ and ${\cal A}$.  From the middle panel of
Fig.~\ref{fig-lptsA} we can see that, for a given value of ${\cal A}$,
$X_{L1}$ increases with increasing values of $q$. For a given value of
$q$, on the other hand, $X_{L1}$ decreases with increasing values of
${\cal A}$.  Additionally, $X_{L1}$ asymptotes to a constant value for
both large and small $q$. This behavior can be understood from
Eq.~(\ref{eq-L123}), which, for small $q$, reduces to
\begin{equation}
X_{L1} \left[ {\cal A} \left(X_{L1} - 1\right)^2  - 
\left(X_{L1} - 1\right) + 1\right] \approx 0.
\label{eq-sqL1}
\end{equation}
The only real solution to this equation for which $0 < X_{L1} < 1$ is
$X_{L1} \approx 0$. Thus, in the limit of small $q$, $X_{L1} \rightarrow
0$.  For large $q$, Eq.~(\ref{eq-L123}) yields
\begin{equation}
X_{L1} \approx {\cal A}^{-1/3}.  \label{XL1largeq}
\end{equation}
This approximation breaks down for ${\cal A} \la 1$, in which case
$X_{L1} \rightarrow 1$ for large values of $q$. This change in 
the behavior of $X_{L1}$ for large values of $q$ is due to the fact 
that the gravitational force exerted by star~2 is no longer negligible 
as $X_{L1} \rightarrow 1$.

For $L_2$, we can see from the top panel of Fig.~\ref{fig-lptsA} that,
for a given value of ${\cal A}$, $X_{L2}$ decreases with increasing
values of $q$, and, for a given value of $q$, $X_{L2}$ decreases with
increasing values of ${\cal A}$. For small and large values of $q$,
$X_{L2}$ furthermore asymptotes to a constant value depending on the
magnitude of ${\cal A}$. This may be understood from Eq.~(\ref{eq-L123})
which, for small $q$, reduces to
\begin{equation}
X_{L2}^3 - \frac{2\,{\cal A}+1}{\cal A}\,X_{L2}^2 
  + \frac{{\cal A}+2}{\cal A}\,X_{L2} - \frac{2}{\cal A} \approx 0.
\end{equation}
This equation admits of one real solution for $X_{L2}$ which is a
function of ${\cal A}$. The analytical expression for the solution is,
however, rather intricate and does not contribute further to
understanding the behavior of $X_{L2}$ for small values of $q$ other
than confirming the dependence of the asymptotic value of $X_{L2}$ on
${\cal A}$.  For large values of $q$, Eq.~(\ref{eq-L123}) yields
\begin{equation}
X_{L2} \approx {\cal A}^{-1/3}.
\end{equation}
This approximation breaks down for ${\cal A} \ga 1$ for the same reason
that Eq.~(\ref{XL1largeq}) breaks down for ${\cal A} \la 1$, i.e., the
gravitational force exerted by star~2 is no longer negligible when
$X_{L2} \rightarrow 1$.  Instead $X_{L2} \approx 1$ for large values of
$q$ and ${\cal A} \ga 1$.

In the case of $L_3$, we can see from the bottom panel of
Fig.~\ref{fig-lptsA} that, for a given value of ${\cal A}$, $X_{L3}$
decreases with increasing values of $q$, while, for a given value of
$q$, $X_{L3}$ increases with increasing values of ${\cal A}$. In
addition, $X_{L3}$ asymptotes to zero for small values of $q$, and to a
constant depending on ${\cal A}$ for large values of $q$. This may again
be understood from Eq.~(\ref{eq-L123}) which, for small values of $q$,
reduces to
\begin{equation}
X_{L3} \left[ {\cal A}\,X_{L3}^2 - (2\,{\cal A}+1) X_{L3} 
  + {\cal A} + 2 \right] \approx 0.
\end{equation}
The only real solution to this equation with $X_{L3}<0$ is $X_{L3}
\approx 0$. For large values of $q$, on the other hand,
Eq.~(\ref{eq-L123}) yields
\begin{equation}
X_{L3} \approx -{\cal A}^{-1/3}.
\end{equation}
for all values of ${\cal A}$.

{\em (ii) $Y_D \ne 0$.} For $Y_D \ne 0$ and $Z_D=0$,
Eqs.~(\ref{eq-gradvx}) and~(\ref{eq-gradvy}) yield two non-linear,
algebraic equations:
\begin{equation}
\label{eq-gradvx2}
\frac{1}{\left(r_D^2-2X_D+1\right)^{3/2}} -1 = 0,
\end{equation}
\begin{equation}
\label{eq-gradvy2}
\frac{q}{r_D^3} +\frac{1}{\left(r_D^2-2X_D+1\right)^{3/2}} - (1+q){\cal 
A}(f,e,\nu) = 0.
\end{equation}
These equations have two real, analytical solutions for $X_D$ and $Y_D$ 
given by
\begin{eqnarray}
\label{eq-L45x}
X_D &=& \frac{1}{2}\left[\frac{q}{(1+q){\cal A}(f,e,\nu)-1}\right]^{2/3},\\
\label{eq-L45y}
 Y_D &=& \pm\sqrt{X_D(2-X_D)}.
\end{eqnarray}
The solutions correspond to two stationary points located on the tips of
triangles whose bases coincide with the line connecting the mass centers
of the binary component stars.  In accordance with the nomenclature for
circular synchronized binaries, we refer to these points as the
triangular Lagrangian points $L_4$ and $L_5$.  The $L_4$ point is the stationary
point associated with $Y_D > 0$, while the $L_5$ point is the stationary
point associated with $Y_D < 0$. Real solutions to Eqs.~(\ref{eq-L45x})
and (\ref{eq-L45y}) furthermore only exist when
\begin{equation}
q > \frac{1-{\cal A}}{{\cal A}-1/8}.
\label{eq-limA1}
\end{equation}
This condition is always satisfied when ${\cal A} \ge 1$.  In the
limiting case where $e=0$ and $f=1$, Eqs.~(\ref{eq-L45x})
and~(\ref{eq-L45y}) reduce to the usual formulae for the position of the
$L_4$ and $L_5$ points in the case of a circular, synchronized binary,
i.e. $X_D = 1/2$ and $Y_D = \pm \sqrt{3}/2$
\citep[e.g.][]{2000ssd..book.....M}.
\clearpage
\begin{figure}
\plotone{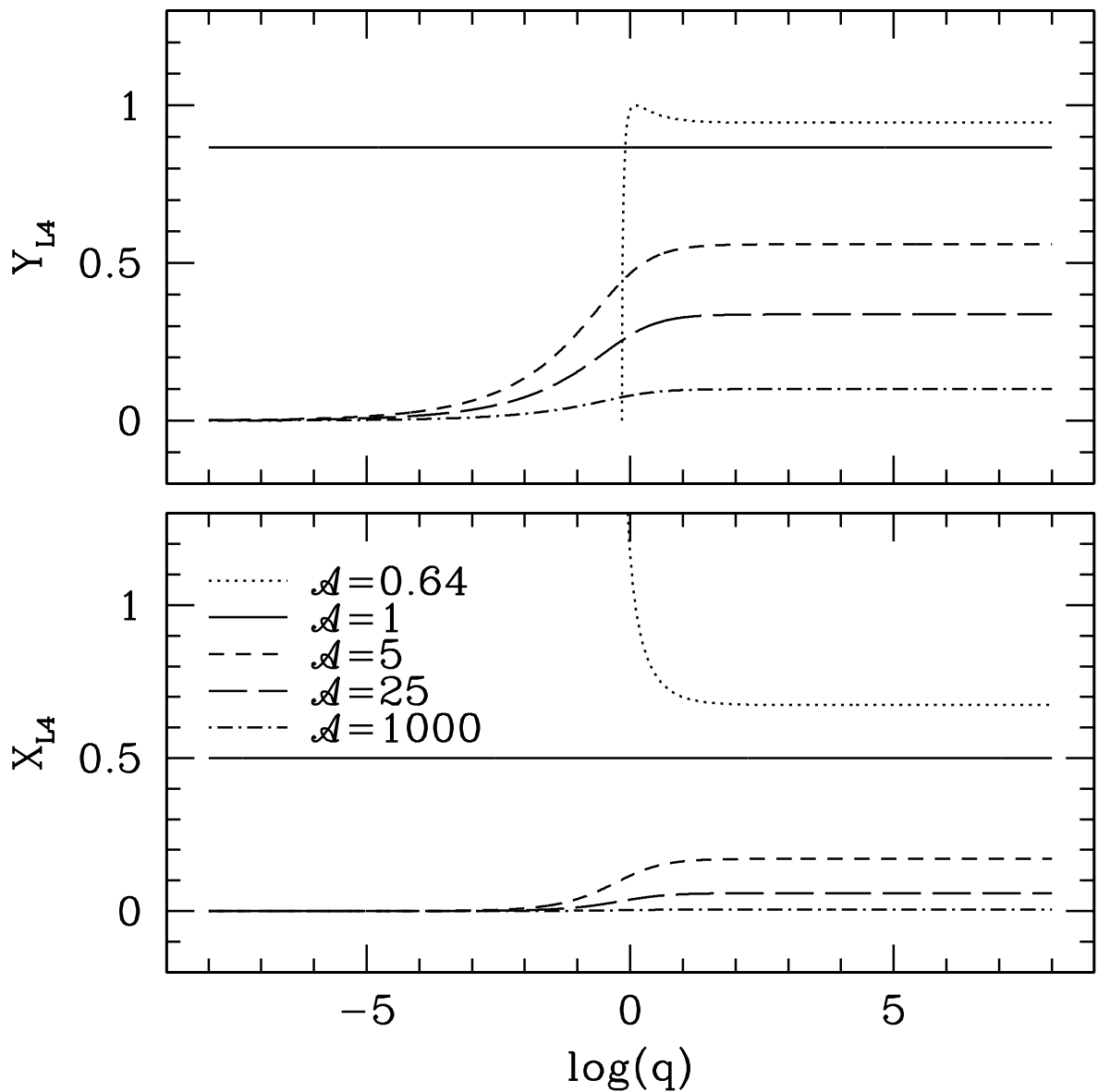}
\caption{The $X_D$ and $Y_D$ coordinates of the leading triangular
Lagrangian point, $L_4$, as a function of $q$ for different values of
${\cal A}$. The dotted line terminates near $q=0.69$ below which $L_4$
does not exist for ${\cal A}=0.64$. The $X_D$ and $Y_D$ coordinates of the
trailing Lagrangian point, $L_5$, is identical to that of $L_4$, except
that $Y_D$ has the opposite sign for $L_5$ as for $L_4$.}
\label{fig-L45A}
\end{figure}
\clearpage
The dependence of the $X_D$ and $Y_D$ coordinates of $L_4$ on $q$
and ${\cal A}$ is illustrated in the bottom and top panels of
Fig.~\ref{fig-L45A}, respectively. Since $X_{L5}=X_{L4}$ and
$Y_{L5}=-Y_{L4}$, the dependence of the position of $L_5$ on $q$ and
${\cal A}$ is similar to that of $L_4$. For a given value of ${\cal A} >
1$, both $X_{L4}$ and $Y_{L4}$ increase with increasing values of $q$.
The coordinates furthermore asymptote to constant values for both small
and large values of $q$. The asymptotic behavior can be understood from
Eq.~(\ref{eq-L45x}) which reduces to
\begin{equation}
X_{L4} \approx 0.5\left( \frac{q}{{\cal A}-1} \right)^{2/3}
\end{equation}
for small values of $q$, and 
\begin{equation}
X_{L4} \approx 0.5\,{\cal A}^{-2/3}  \label{XL4largeq}
\end{equation}
for large values of $q$. Thus, in the limit of small $q$, $X_{L4}
\rightarrow 0$ when ${\cal A} > 1$. For ${\cal A}<1$, the $L_4$ point no
longer exists for all values of $q$. E.g., when ${\cal A} = 0.64$ (the
dotted line in Fig.~\ref{fig-L45A}), the $L_4$ point only exists for $q
> 0.69$ (see Eq.~\ref{eq-limA1}). However, when the $L_4$ point exists,
$X_{L4}$ still asymptotes to the value given by Eq.~(\ref{XL4largeq})
for large values of $q$. Finally, for a given value of $q$, $X_{L4}$ and
$Y_{L4}$ decrease with increasing values of ${\cal A}$. In addition, it
is interesting to note that for ${\cal A}>1$, $L_4$ and $L_5$ are always
located inside the binary orbit, while for ${\cal A} < 1$, $L_4$ and
$L_5$ are located outside the binary orbit.

\subsection{Stability}
\label{sec-stability}
\clearpage
\begin{figure*}
\plottwo{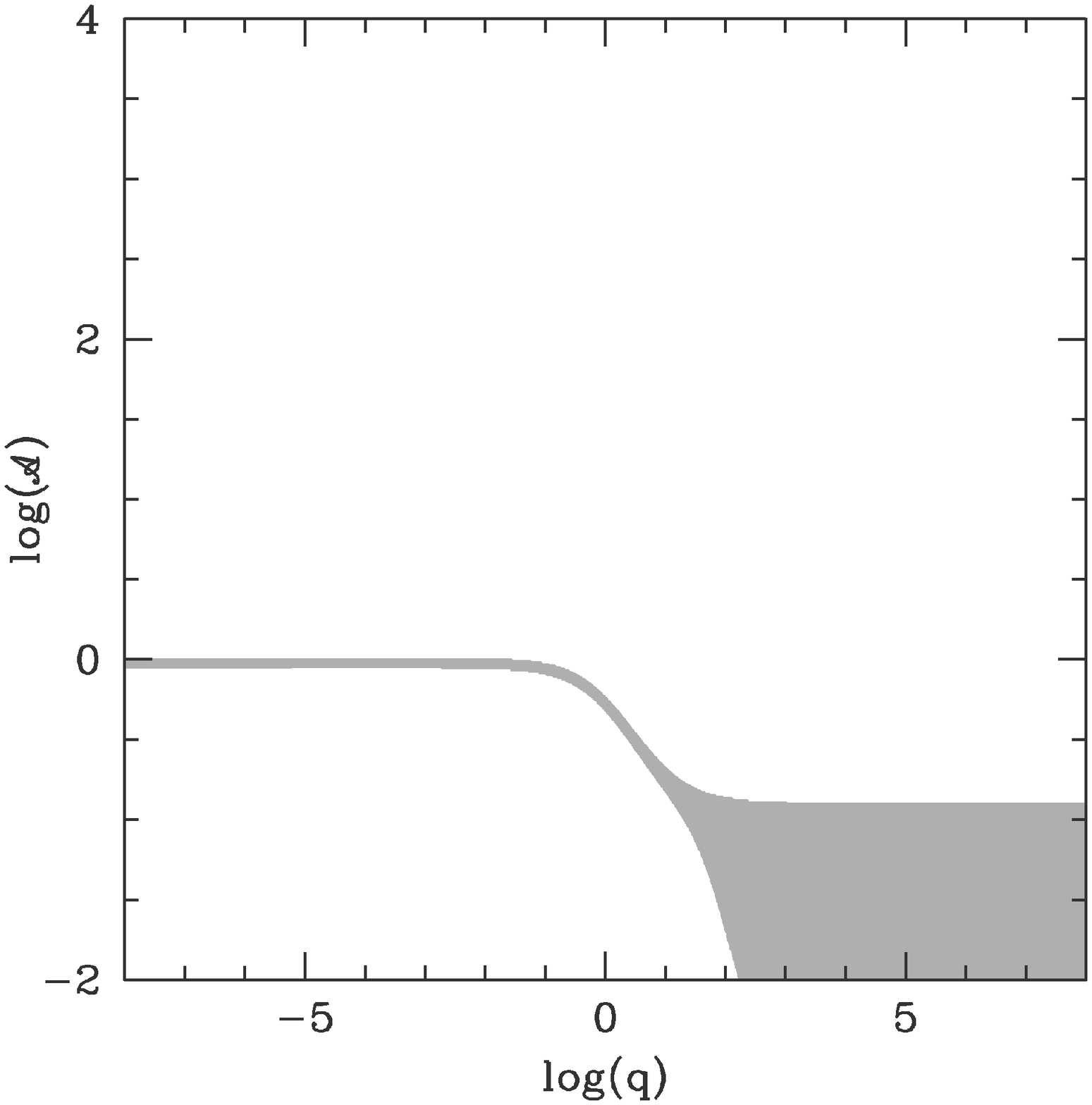}{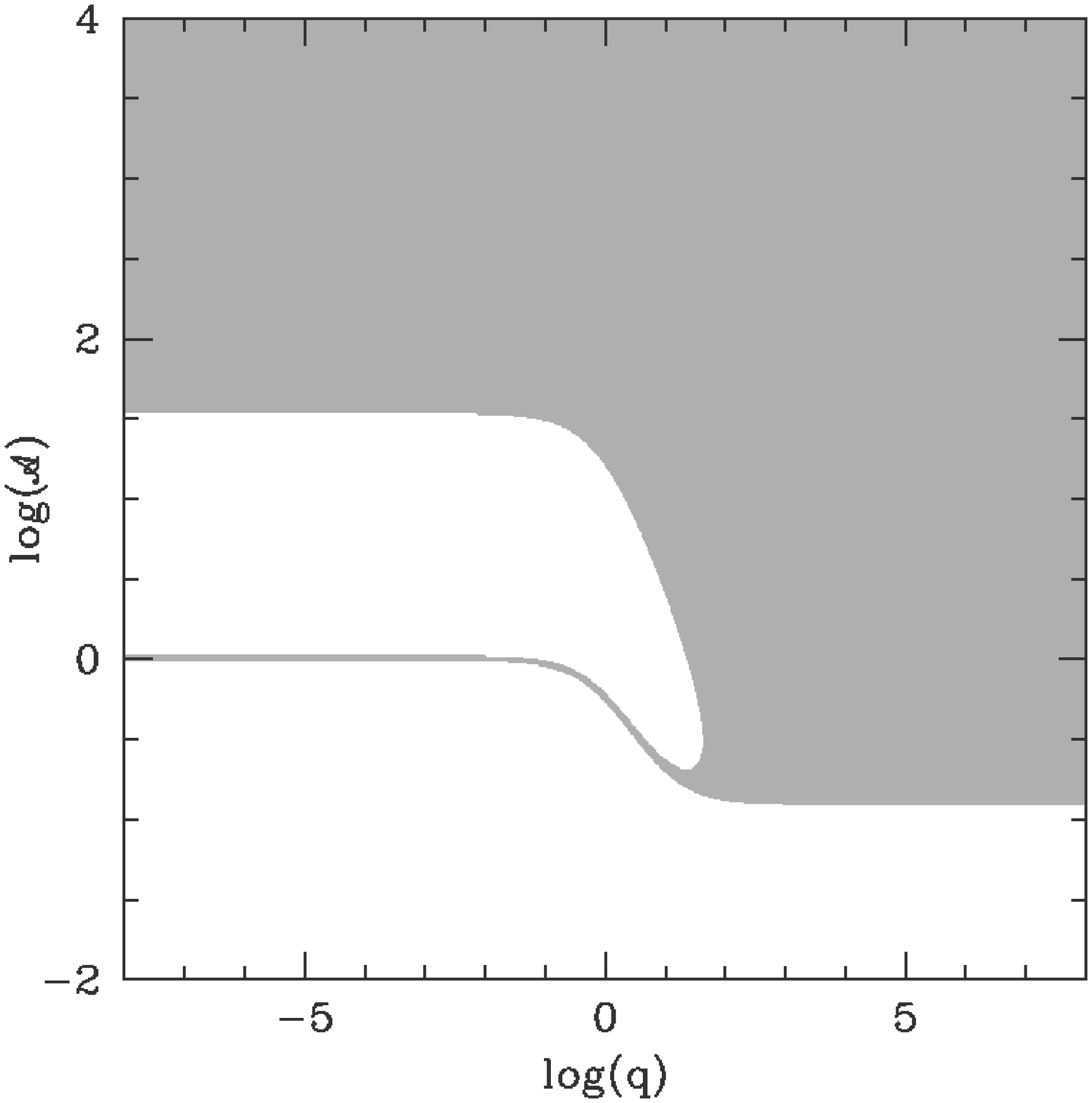}
\caption{ Stability regions (gray-shaded) for the Lagrangian points
$L_2$ (left) and $L_4$, $L_5$ (right) in the $(q,{\cal A})$-plane. The
$L_2$ stability region remains always at $\log{\cal A} < 0$, in
agreement with the known results for circular, synchronous binaries
($\log{\cal A} = 0$) for which $L_2$ is always unstable.  The stability
regions for $L_4$ and $L_5$ are identical to each other since the
stability conditions given by
Eqs.~(\ref{eq-stacond1})--(\ref{eq-stacond3}) are dependent only on the
square of the $Y_D$ coordinate.}
\label{fig-stability}
\end{figure*}
\clearpage
Following the methodology described by \citet{2000ssd..book.....M}
(Chapter~3.7) for circular, synchronous binaries, we examine the
stability of the Lagrangian points in eccentric, non-synchronous
systems.  We employ a linear stability analysis where we perturb
and linearize the equation governing the motion of a mass element in the
binary potential and follow the behavior of an element given a small
displacement from one of the Lagrangian points. If the solution to the
perturbed equation of motion is oscillatory, the Lagrange point is
stable to small perturbations. All other solutions imply instability. 
For stability, then, it follows that all three of the following 
conditions on the second derivatives of the dimensionless potential 
$V_D$ must be satisfied.  These are
\begin{eqnarray}
\label{eq-stacond1}
B^2 &>& 4C, \\
\label{eq-stacond2}
\sqrt{B^2-4C} &\leq& B, \\
\label{eq-stacond3}
\sqrt{B^2-4C} &\geq& -B, 
\end{eqnarray}
where
\begin{eqnarray}
\label{eq-BC}
B &\equiv& 4{\cal A}(1+q) +V_{XX} + V_{YY}, \\
C &\equiv& V_{XX}V_{YY} - V_{XY}^2,
\end{eqnarray}
$V_{XX} \equiv \partial^2V_D/\partial X_D^2 $, $V_{YY} \equiv
\partial^2V_D/\partial Y_D^2$, and $V_{XY}\equiv\partial^2V_D/\partial
X_D \partial Y_D$ are the second derivatives of the potential
$V_D$ with respect to $X_D$ and $Y_D$, and the square roots in 
Eqs. (\ref{eq-stacond2}) and (\ref{eq-stacond3}) are assumed to be 
positive.  Examining these conditions, we
find that $L_1$ and $L_3$ are always unstable for any combination of
binary parameters, while $L_2$, $L_4$, and $L_5$ can be stable as well
as unstable depending on the values of $q$ and ${\cal A}$. The regions
in the $(q,{\cal A})$-plane where $L_2$, $L_4$, and $L_5$ are stable 
and are shown by the gray-shaded areas in Fig.~\ref{fig-stability}.

For $q \la 0.1$, the $L_2$ point is found to be stable for a narrow
range of ${\cal A}$-values very near, but {\it always less than} ${\cal
A} = 1$, while for $q \ga 10$, a broad stable region is found at ${\cal
A} \la 0.1$ (Fig.~\ref{fig-stability}, left panel). In the latter
regime, the potential is very similar to that of a non-rotating single
star, except for a small contribution due to the gravitational potential
of star~2. Even though this contribution is small, it is significant
enough to be the sole cause of this stable region.  This is further
supported by the fact that, in this regime, the potential at $L_3$ is
identical to that at $L_2$ except for this small contribution, and yet
$L_3$ is never stable throughout the considered parameter space.  Thus,
the second body, however small, is the sole cause for the stable regions
of $L_2$.  We note that we do not find any stable points for $\log{\cal
A}=0$, in agreement with the analysis by \citet{2000ssd..book.....M} for
circular, synchronous binary systems.

The stability region for the triangular Lagrangian points, $L_4$ and
$L_5$, covers a wide range of $q$ and ${\cal A}$ values which, for $q
\la 0.1$, converges into a narrow stability region near ${\cal A}=1$
(Fig.~\ref{fig-stability}, right panel).  Solutions at exactly ${\cal
A}=1$ exist only for $\log{q} \leq -1.4$ and $\log{q} \geq 1.4$, in
agreement with the analysis of \citet{2000ssd..book.....M} for circular,
synchronized binaries (see their Eq.~3.145). It is apparent, that the
$L_4$ and $L_5$ stability regions are much larger for eccentric,
non-synchronous systems than for circular, synchronous systems. The
existence of these stable points is important to study the trapping of
test particles such as gas attempting to escape the system or even
Trojan asteroids in planetary systems \citep[see][for a discussion on
Trojan asteroids around extrasolar planets]{2006astro.ph..9298F}.  It
should also be noted that Lagrangian points within the stable regions
shown in Fig.~\ref{fig-stability} need not remain stable throughout the
binary orbit due to the dependence of ${\cal A}$ on the mean anomaly
(see Eq.~\ref{eq-A} and Fig.~\ref{fig-A}).

\section{The Volume-Equivalent Roche-Lobe Radius}
\label{sec-veqrlr}

Under the quasi-static approximation, the maximum size of a star in a
close binary is determined by the equipotential surface connected to the
inner Lagrangian point $L_1$.  This surface is commonly referred to as
the Roche lobe. Once a star fills its Roche lobe, any further expansion
of the star or shrinkage of the lobe results in mass loss from the star. 
For a circular binary with synchronously rotating component stars, the
shape and volume of the Roche lobe depend solely on the mass ratio, $q$,
of the system.  For eccentric binaries with non-synchronous component
stars, the shape and volume of the Roche lobe also depend upon the
eccentricity, the true anomaly, and the degree of asynchronism. As
before, we group these dependencies into the parameter ${\cal
A}(f,e,\nu)$ (see Eq.~\ref{eq-A}). \clearpage \begin{figure*}
\plotone{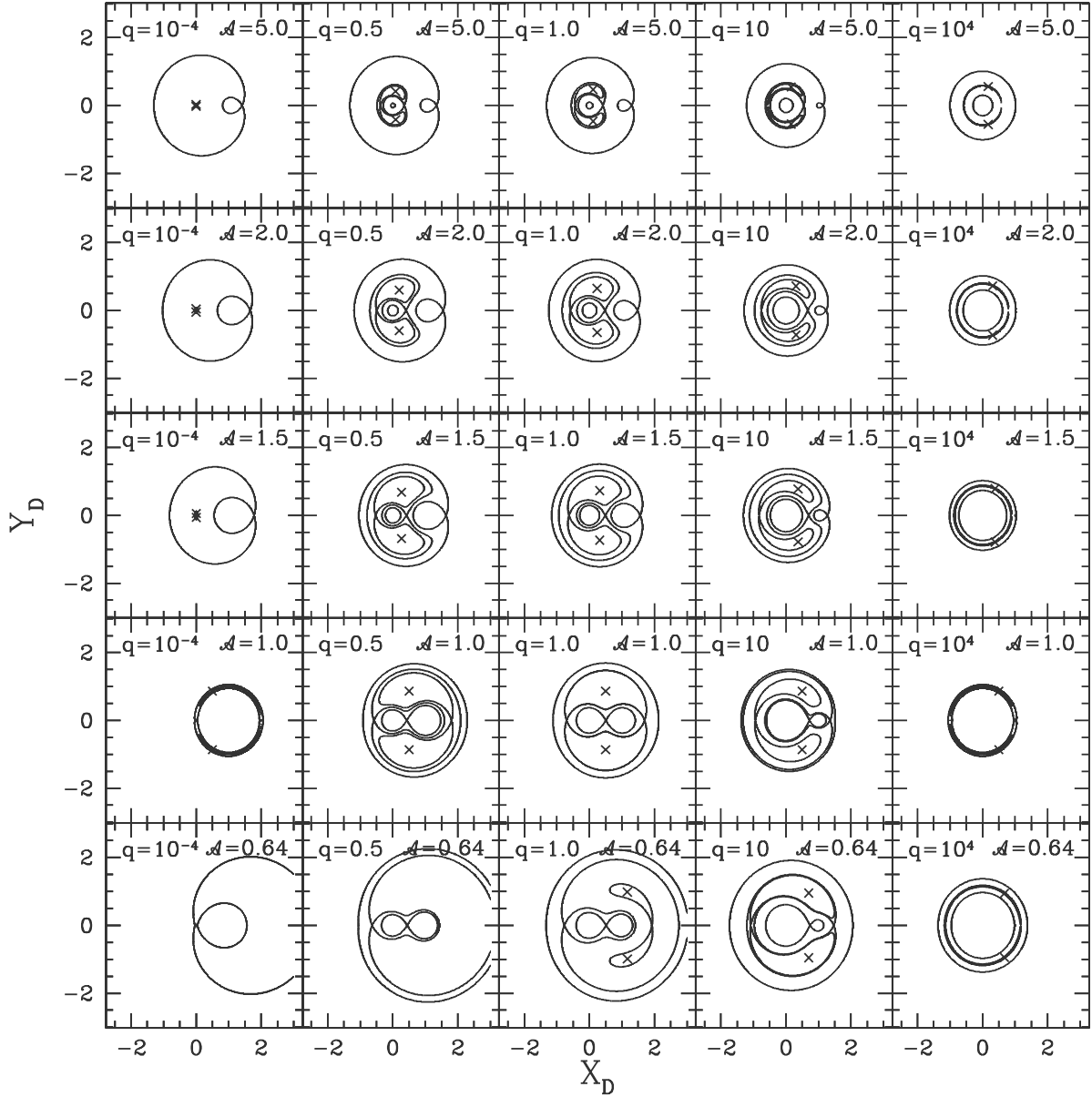} \caption{Instantaneous equipotential surfaces of $V_D$
given by Eq.~(\ref{eq-VD}) in the $Z=0$ plane for different mass ratios
$q=M_1/M_2$ and values of the parameter ${\cal A}(f, e, \nu)$. The
contours shown correspond to the equipotential surfaces passing through
the co-linear Lagrangian points $L_1$, $L_2$, and $L_3$.  The locations
of the triangular Lagrangian points $L_4$ and $L_5$, if they exist, are
marked with crosses.  Star~1 is located at $(X_D,Y_D)=(0,0)$, and star~2
at $(X_D,Y_D)=(1,0)$. The ${\cal A}=1.0$ panels correspond to circular
binaries with synchronously rotating component stars. For ${\cal
A}=0.64$, $L_4$ and $L_5$ only exist for $q > 0.69$ (see
Eq.~\ref{eq-limA1}).} \label{fig-equipots} \end{figure*} \clearpage The
variations in the shape and size of the binary equipotential surfaces
are shown in Fig.~\ref{fig-equipots} as functions of $q$ and ${\cal A}$.
The contours shown correspond to the cross-sections of the equipotential
surfaces passing through the co-linear Lagrangian points with the
$Z_D=0$ plane. The locations of the triangular Lagrangian points, if
they exist, are marked with crosses.  For large mass ratios, the
equipotential surfaces resemble that of a rotating single star
regardless of the value of ${\cal A}$. For smaller mass ratios, the
structure of the equipotential surfaces becomes more complex and the
detailed shapes become more sensitive to the value of ${\cal A}$.  A
particularly interesting observation is that for ${\cal A} > 1$ the
equipotential surface passing through $L_1$ may ``open up'' and no
longer enclose star~2\footnote{Note that this ``opening up'' of the
Roche potential has also been seen in studies of radiation effects on
the shape of the equipotential surfaces.  See
\citet{1972Ap&SS..19..351S} and \citet{1977A&A....54..877V}.}.  This
geometry may facilitate mass loss from the system through $L_2$ when
star~1 fills its Roche lobe and transfers mass to star~2 through $L_1$.
We will discuss this in more detail in the next section. For values of
${\cal A} \ga 10$, the centrifugal term dominates the potential $V_D$,
so that the equipotential surfaces resemble those of a rotating single
star regardless of the mass ratio $q$.

The equipotential surfaces shown in Fig.~\ref{fig-equipots} differ
considerably from those of \citet{2005MNRAS.358..544R}. In particular,
the orbital parameters adopted in Figs. 3 and 4 of that paper yield
${\cal A}$-values of 1.4 and 1.8, respectively, which may be compared to
the ${\cal A}=1.5$ and ${\cal A}=2.0$ panels for $q=0.5$ in
Fig.~\ref{fig-equipots}. The differences may be attributed to the
erroneous use of the \citet{1985ibs..book.....P} potential in
\citet{2005MNRAS.358..544R} which is only valid for binaries with
circular orbits.

In order to determine whether or not a star fills its critical Roche
lobe, a full three-dimensional treatment of stellar and binary evolution
is, in principle, required. Such a treatment is, however, prohibitively
computationally expensive. Roche-lobe overflow is therefore usually
studied using one-dimensional stellar and binary evolution codes in
which the Roche lobe of a star is approximated by a sphere of equivalent
volume. The star of radius ${\cal R}_1$ is then assumed to overflow its
Roche lobe and transfer mass to its companion when ${\cal R}_1 > {\cal
R}_L$, where ${\cal R}_L$ is the radius of the sphere with volume equal
to that of the Roche lobe.
\clearpage
\begin{figure}
\plotone{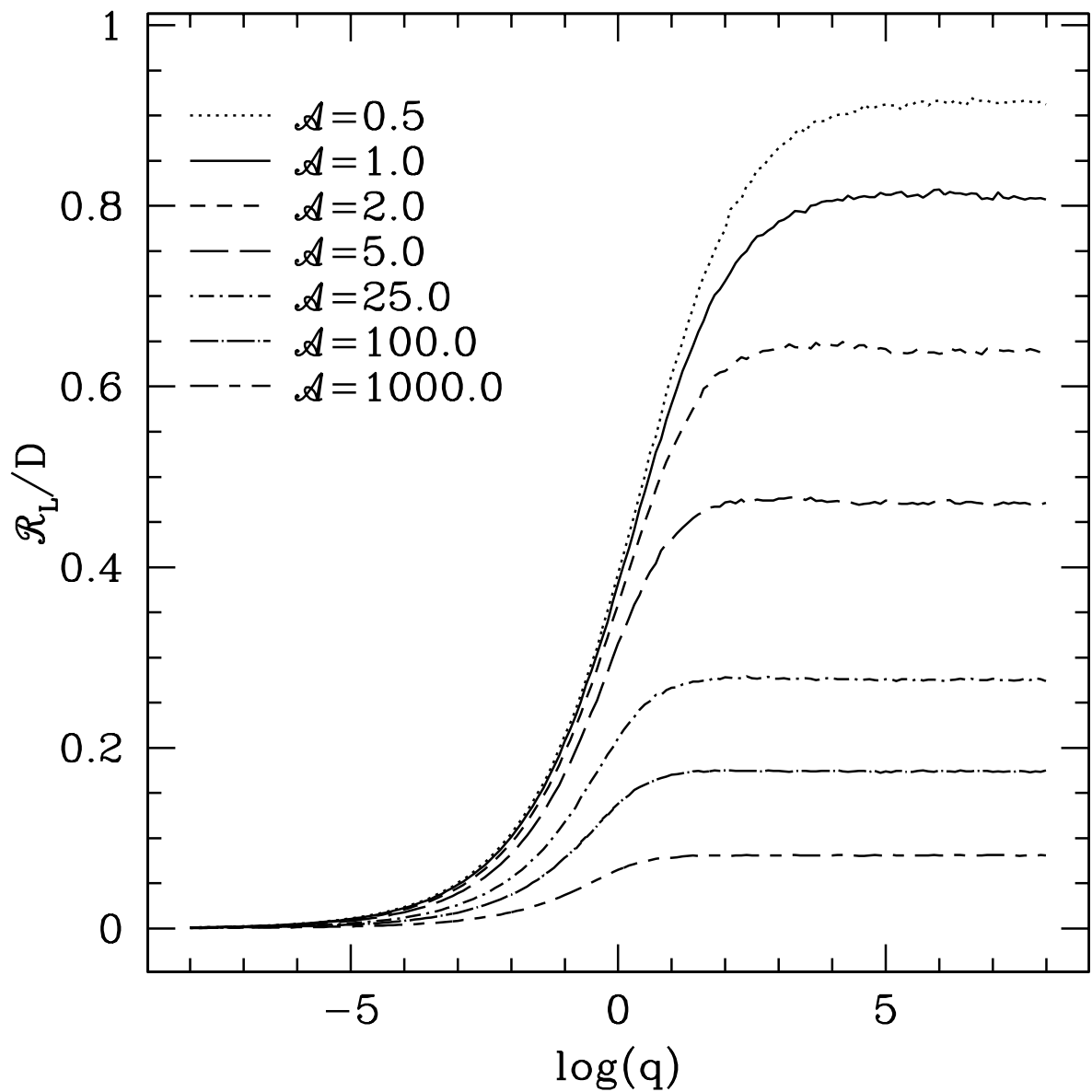}
\caption{The volume-equivalent Roche lobe radius, ${\cal R}_L$ in units 
of the instantaneous distance $D(t)$ between the binary components as a 
function of $\log q$, for a range of ${\cal A}$-values.  Small 
fluctuations in the curves are due to the finite errors in the Monte 
Carlo integration method used to calculate the volume of the Roche 
lobe.}
\label{fig-eqrlr}
\end{figure}
\clearpage
In Fig.~\ref{fig-eqrlr}, we show the volume-equivalent Roche lobe radius
in units of the instantaneous distance between the binary components
calculated via Monte Carlo volume integration as a function of the mass
ratio, $q$, for a range of ${\cal A}$-values.  We note that our
calculation of the Roche lobe radius agrees with that of
\citet{1977A&A....54..877V} to within a few percent in the regime in
which our parameter spaces intersect (circular, non-synchronous orbits
where the effects of radiation pressure have been neglected). 
Comparison of Fig.~\ref{fig-eqrlr} with Fig.~\ref{fig-lptsA} shows that
${\cal R}_L$ as a function of $q$ is similar in shape to $L_1$ as a
function of $q$, which is expected since it is the potential at $L_1$
which defines the size and shape of the Roche lobe.  For small values of
$q$, ${\cal R}_L$ asymptotes to zero independent of ${\cal A}$ because
the location of $L_1$ asymptotes to the position of star~1. For large
values of $q$, ${\cal R}_L$ asymptotes to an ${\cal A}$-dependent value.

A simple and accurate fitting formula for the volume-equivalent Roche
Lobe radius of a star in a circular binary with synchronously rotating
component stars has been provided by \citet{1983ApJ...268..368E}:
\begin{equation}
{\cal R}_{L,{\rm circ}}^{\rm Egg} = a\,\frac{0.49\,q^{2/3}}{0.6\,q^{2/3} 
  + \ln \left( 1 + q^{1/3} \right)},
\end{equation}
where $a$ is the radius of the circular orbit. For lack of a better
treatment, this formula is often extrapolated to binaries with eccentric
orbits. In particular, Roche lobe radii of stars at the periastron of a
binary orbit are commonly approximated by
\begin{equation}
{\cal R}_{L,{\rm peri}}^{\rm Egg} = a(1-e)\,\frac{0.49\,q^{2/3}}{0.6\,q^{2/3} 
  + \ln \left( 1 + q^{1/3} \right)},
\label{eq-rleggAperi}
\end{equation}
where $a(1-e)$ is the periastron distance of the binary orbit.  We can
further extend this generalization and apply it to arbitrary orbital 
phases by replacing the periastron distance between the stars with the 
current orbital separation, $D(t)$.  Thus,
\begin{equation}
{\cal R}_{L}^{\rm Egg} = D(t)\,\frac{0.49\,q^{2/3}}{0.6\,q^{2/3} 
  + \ln \left( 1 + q^{1/3} \right)}.
\label{eq-rleggA}
\end{equation}
\clearpage
\begin{figure}
\plotone{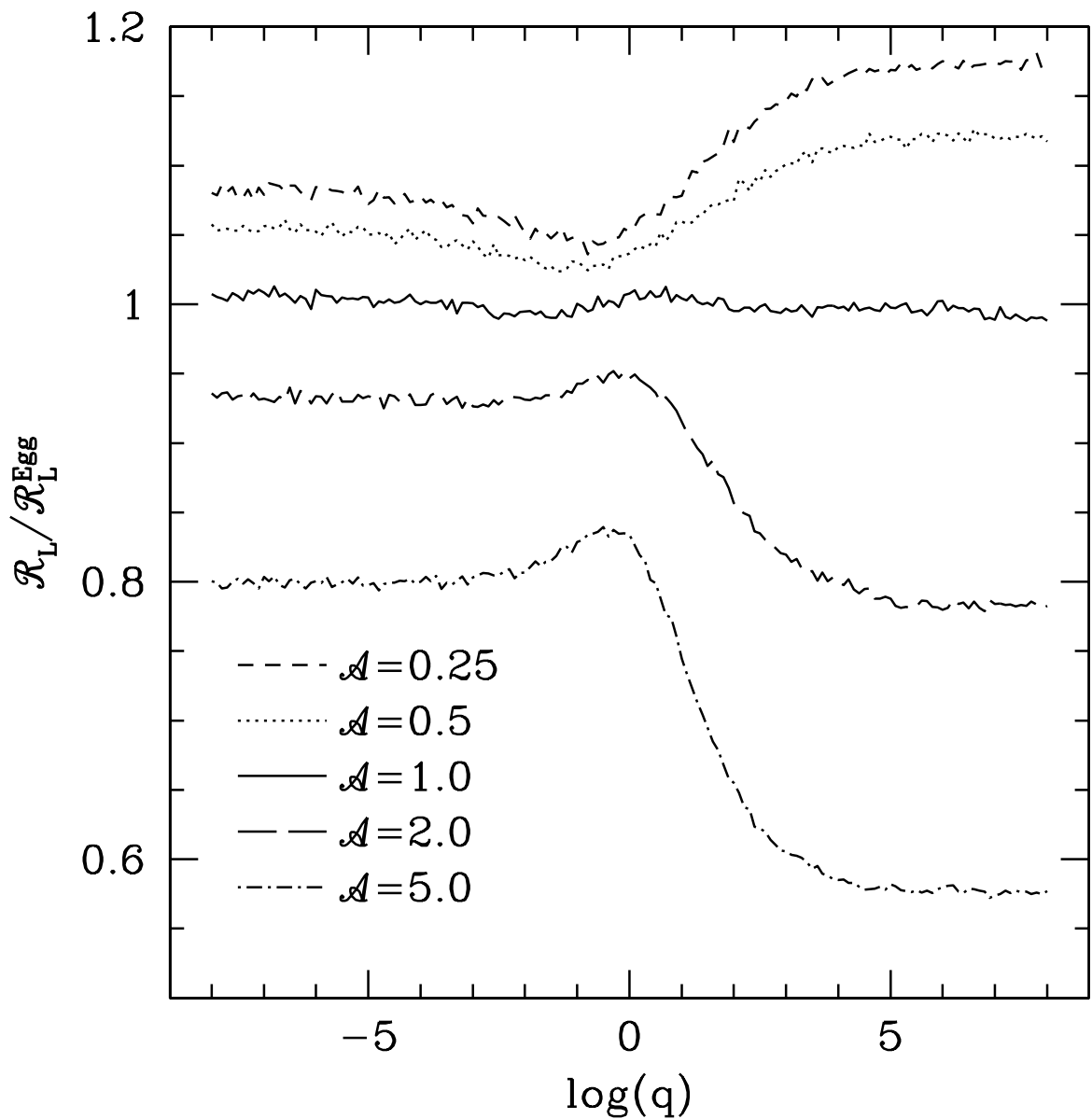}
\caption{The ratio of the volume-equivalent Roche Lobe radius at 
periastron obtained by numerically calculating the volume of the Roche 
lobe to the volume-equivalent Roche Lobe radius obtained from 
Eq.~(\ref{eq-rleggA}).  Small fluctuations in the curves are due to the 
finite errors in the Monte Carlo integration method used to calculate 
the volume of the Roche lobe.}
\label{fig-RLRLegg}
\end{figure}
\clearpage
In Fig.~\ref{fig-RLRLegg}, we show the ratio of the
volume-equivalent Roche Lobe radius at periastron calculated through a
Monte Carlo integration method to the volume-equivalent Roche Lobe
radius determined by means of Eq.~(\ref{eq-rleggA}). For ${\cal A}
\approx 1$, we find Eq.~(\ref{eq-rleggA}) to be accurate to within a few
percent over a wide range of $q$-values, confirming previous statements
on the accuracy of the equation based on smoothed particle hydrodynamics
calculations \citep{2005Icar..175..248F, 2005MNRAS.358..544R}.  For a
given value of ${\cal A}$, the ratio ${\cal R}_L/{\cal R}_L^{\rm Egg}$
is closest to unity near $q=1$ and indicates better agreement between
${\cal R}_L$ and ${\cal R}_L^{\rm Egg}$ for $q<1$ than for $q>1$,
especially for large values of ${\cal A}$ (by almost a factor of 2 as
shown in Fig.~\ref{fig-RLRLegg}). For a given value of $q$, the
agreement between ${\cal R}_L$ and ${\cal R}_L^{\rm Egg}$ furthermore
becomes progressively worse as ${\cal A}$ differs from unity.

Hence, a star rotating synchronously with the orbital angular velocity
at periastron in an eccentric binary (so that $1 \lesssim {\cal A}
\lesssim 2$ for $e < 0.9$) has a smaller volume-equivalent Roche Lobe
radius at periastron than a star rotating synchronously in a circular
orbit with radius equal to the periastron distance of the eccentric
binary. As shown in Fig.~\ref{fig-RLRLegg}, this decrease in ${\cal
R}_L$ may be substantially smaller than the $(1-e)$ factor introduced in
Eq.~(\ref{eq-rleggAperi}). Consequently, the maximum size of star~1 is
smaller in an eccentric binary and mass transfer from star~1 to star~2
may begin earlier than in a circular binary with the same instantaneous
distance between the component stars.

For ease of use, we provide a fitting formula for ${\cal R}_L/{\cal 
R}_L^{\rm Egg}$ as a function of $q$ and ${\cal A}$.  In order to keep 
the accuracy of the fit better than 1\%, it is necessary to divide the 
parameter space into six different regimes.  The formulae below are 
accurate to better than 1\% for $-8 \leq \log{q} \leq 8 $ and $-2\leq 
\log{\cal A} \leq 4$, and, in most cases, the accuracy is better than 
0.5\%.

\begin{eqnarray}
\frac{{\cal R}_L}{{\cal R}_L^{\rm Egg}}&&[\log{q} \geq 0; \log{\cal A}
\leq -0.1 ]=\nonumber \\
&& 1.226 -0.21 {\cal A} - 0.15(1-{\cal A})\exp[(0.25{\cal A} 
-0.3)(\log{q})^{1.55}]
\end{eqnarray}

\begin{eqnarray}
\frac{{\cal R}_L}{{\cal R}_L^{\rm Egg}}&&[\log{q} \leq 0; \log{\cal A}
\leq -0.1 ]=\nonumber \\
&&1 + 0.11(1-{\cal A}) - 0.05(1-{\cal A})\exp[-(0.5(1+{\cal 
A})+\log{q})^2]
\end{eqnarray}

\begin{eqnarray}
\frac{{\cal R}_L}{{\cal R}_L^{\rm Egg}}&&[\log{q} \leq 0; -0.1 \geq 
\log{\cal A}\leq 0.2 ]= g_0({\cal A}) + g_1({\cal A})\log{q} + g_2({\cal 
A})(\log{q})^2 \\
&&g_0({\cal A}) = 0.9978 - 0.1229\log{\cal A} -0.1273(\log{\cal A})^2 
\nonumber \\
&&g_1({\cal A}) = 0.001 + 0.02556\log{\cal A} \nonumber \\
&&g_2({\cal A}) = 0.0004 + 0.0021\log{\cal A} \nonumber
\end{eqnarray}

\begin{eqnarray}
\frac{{\cal R}_L}{{\cal R}_L^{\rm Egg}}&&[\log{q} \geq 0; -0.1 \geq 
\log{\cal A}\leq 0.2 ]= h_0({\cal A}) + h_1({\cal A})\log{q} + h_2({\cal 
A})(\log{q})^2 \\
&&h_0({\cal A}) = 1.0071 - 0.0907\log{\cal A} -0.0495(\log{\cal A})^2 
\nonumber \\
&&h_1({\cal A}) = -0.004 - 0.163\log{\cal A} -0.214(\log{\cal A})^2 
\nonumber \\
&&h_2({\cal A}) = 0.00022 - 0.0108\log{\cal A} -0.02718(\log{\cal A})^2 
\nonumber
\end{eqnarray}

\begin{eqnarray}
\frac{{\cal R}_L}{{\cal R}_L^{\rm Egg}}&&[\log{q} \leq 0; \log{\cal 
A} \geq 0.2 ]= i_0({\cal A}) + i_1({\cal A})\exp\left[-i_2({\cal 
A})(\log{q} + i_3({\cal A}))^2 \right] \\
&&i_0({\cal A})=\frac{6.3014(\log{\cal 
A})^{1.3643}}{\exp[2.3644(\log{\cal A})^{0.70748}] - 
1.4413\exp[-0.0000184(\log{\cal A})^{-4.5693}]} \nonumber \\
&&i_1({\cal A})=\frac{\log{\cal A}}{0.0015\exp[8.84(\log{A})^{0.282}] + 
15.78} \nonumber \\
&&i_2({\cal A})=\frac{1 + 0.036\exp[8.01(\log{\cal 
A})^{0.879}]}{0.105\exp[7.91(\log{\cal A})^{0.879}]} \nonumber \\
&&i_3({\cal A})=\frac{0.991}{1.38\exp[-0.035(\log{\cal A})^{0.76}] + 
23.0\exp[-2.89(\log{\cal A})^{0.76}]} \nonumber
\end{eqnarray}

\begin{eqnarray}
\frac{{\cal R}_L}{{\cal R}_L^{\rm Egg}}&&[\log{q} \geq 0; \log{\cal
A} \geq 0.2 ]= j_0({\cal A}) + j_1({\cal A})\exp\left[-j_2({\cal
A})(\log{q})^{j_3({\cal A})} \right] \\
&&j_0=\frac{1.895(\log{\cal A})^{0.837}}{\exp[1.636(\log{A})^{0.789}]-1} 
\nonumber \\
&&j_1=\frac{4.3(\log{\cal A})^{0.98}}{\exp[2.5(\log{\cal 
A})^{0.66}]+4.7} \nonumber \\
&&j_2=\frac{1}{8.8\exp[-2.95(\log{\cal 
A})^{0.76}]+1.64\exp[-0.03(\log{\cal A})^{0.76}]} \nonumber \\
&&j_3=0.256\exp[-1.33(\log{\cal A})^{2.9}](5.5\exp[1.33(\log{\cal 
A})^{2.9}]+1) \nonumber
\end{eqnarray}

\section{Systemic mass loss}
\label{sec-massloss}

As mentioned in the previous section, once star~1 fills its Roche lobe,
any further stellar expansion or orbital contraction may cause the star
to begin losing mass through $L_1$. In the standard case of a circular
synchronous binary, any matter flowing from star~1 through $L_1$ will,
in the absence of other non-gravitational or rotational effects, be
captured by star~2, either forming a disk of material or falling
directly onto its surface. In the case of an eccentric and/or
non-synchronous binary, the geometry of the equipotential surfaces is
such that this is not necessarily the case.
\clearpage
\begin{figure}
\plotone{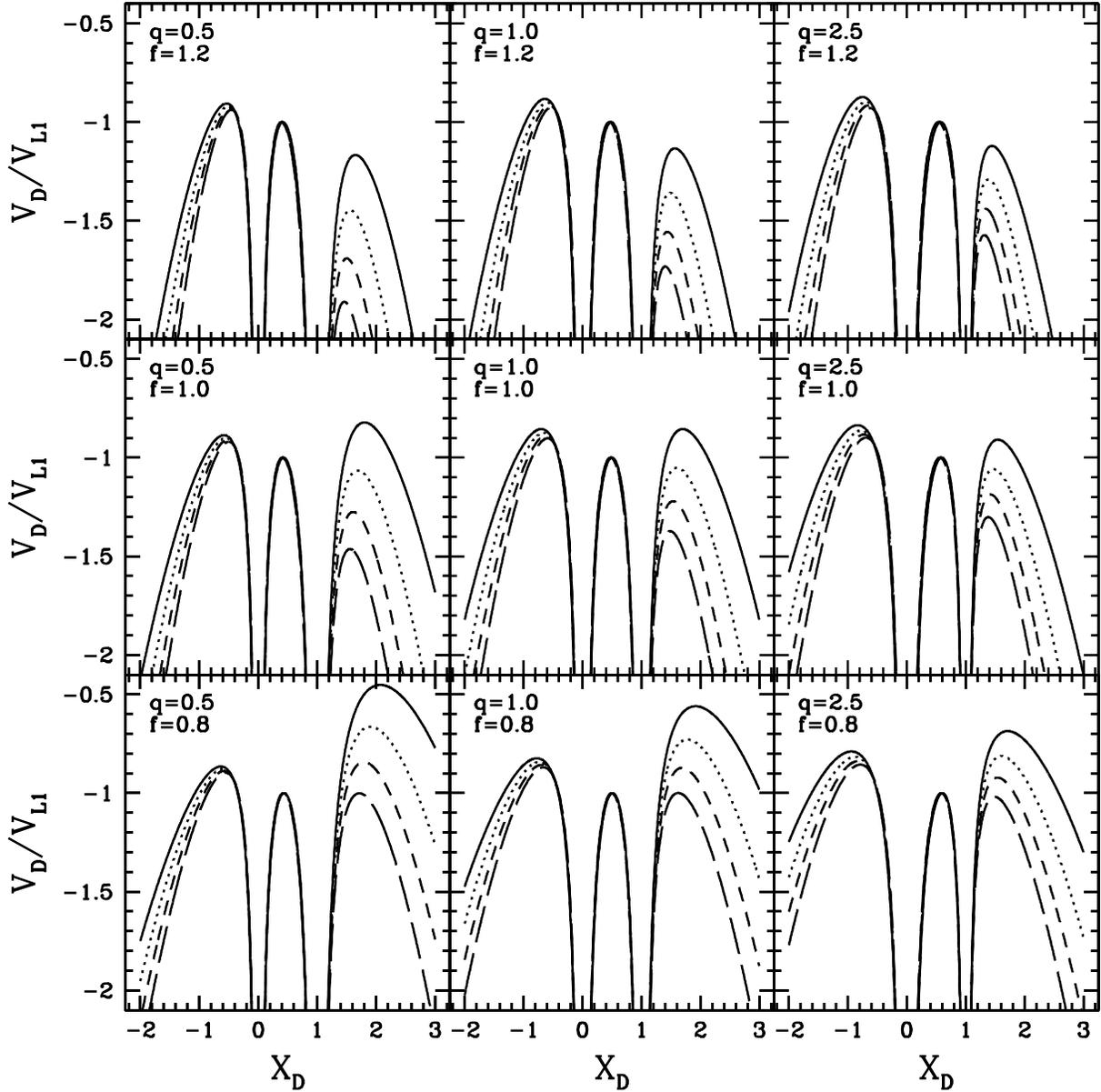}
\caption{Height of the dimensionless potential $V_D$ at periastron
along the $X_D$-axis ($Y_D=Z_D=0$) for different values of the mass
ratio, orbital eccentricity, and ratio of rotational to orbital angular
velocity at periastron. Different line types in each panel correspond to
the orbital eccentricities $e=0$ (solid lines), $e=0.3$ (dotted lines),
$e=0.6$ (short-dashed lines), and $e=0.9$ (long-dashed lines). In each
panel, the maxima in the curves from left to right correspond to the
$L_3$, $L_1$, and $L_2$ point. Star~1 is located at $X_D =0$ and star~2
at $X_D=1$. For ease of comparison, the curves have been normalized to 
the height of the potential at $L_1$.}
\label{fig-mass}
\end{figure}
\clearpage
In particular, we have seen from Fig.~\ref{fig-equipots} that for
significantly large ${\cal A}$, the equipotential surface passing
through $L_1$ no longer encloses star~2. Matter transferred from star~1
to star~2 through the inner Lagrangian point $L_1$ may therefore be lost
from the system. Whether or not this happens depends on the particle
trajectories in the mass-transfer stream which we will address in detail
in a forthcoming paper. For the purpose of this paper, we restrict
ourselves to comparing the height of the potential at the three
co-linear Lagrangian points as an indicator of possible mass loss
scenarios in eccentric binaries. 

Since the first occurrence of mass transfer in eccentric binaries is
expected to take place when the stars are closest to each other, we
focus our discussion on the height of the potential at the periastron of
the binary orbit. In addition, the graphical representation of the
height of the potential turns out to be clearer in terms of the mass
ratio, orbital eccentricity, and ratio of rotational to orbital angular
velocity at periastron than in terms of the parameter ${\cal A}$. Hence,
the height of the potential $V_D$ at periastron along the line
connecting the mass centers of the component stars (the $X_D$-axis) is
shown in Fig.~\ref{fig-mass} as a function of $q$, $e$, and $f$. To
facilitate the comparison of the relative height of the potential at the
co-linear Lagrangian points, the potential is furthermore normalized to
its value at the inner Lagrangian point $L_1$. 

For the considered ranges of mass ratios, eccentricities, and 
rotation rates, the relative height of the potential at $L_2$ depends 
most sensitively on $e$ and $f$, while the relative height of the 
potential at $L_3$ is largely constant over the considered range of 
parameters.  For a constant $f$ and $q$, an increasing eccentricity tends 
to decrease the relative height of the potential.  The decrease is 
stronger for larger $f$-values and smaller $q$-values.

For the considered parameter ranges, the shape of the normalized
potential near $L_1$ is furthermore highly insensitive to the orbital
parameters, while the height of the potential, as well as the position
of the peak, can change for both $L_2$ and $L_3$.  Most interestingly,
the potential at $L_2$ may be higher than, equal to, or lower than the
potential at the $L_1$ point.  Thus, matter flowing from star~1 through
the $L_1$ point will not necessarily be captured by star~2, and may be
energetic enough to escape the system.  As such, it is possible that
mass transfer through $L_1$ can be significantly non-conservative due to
the geometry of the equipotential surfaces in eccentric binaries, but a
calculation of the amount of matter lost from the system will require
detailed particle trajectories, which is beyond the scope of this paper. 
We also note that the value of the potential at $L_3$ can, for certain
values of the orbital parameters, be very near to the value of the
potential at $L_1$.  In cases such as this, it may be possible for a
small amount of matter also to be lost from star~1 through $L_3$ in
addition to that lost through $L_1$.  However, contrary to
\citet{2005MNRAS.358..544R}, we find the potential at $L_3$ to be always
higher than the potential at $L_1$, so that mass loss through $L_3$
never takes place {\it prior} to mass loss through $L_1$\footnote{Note
that \citet{2005MNRAS.358..544R} adopt a nomenclature for $L_2$ and
$L_3$ that is opposite to ours.}.  This difference may be attributed to
the erroneous use of the \citet{1985ibs..book.....P} potential by
\citet{2005MNRAS.358..544R}.


\section{Summary}
\label{sec-conclusion}

In this paper, we present a detailed investigation of the existence and
properties of equipotential surfaces and Lagrangian points in eccentric
binary star and planetary systems with non-synchronously rotating
component stars or planets. The analysis is valid as long as the time
scale of any tidally induced oscillations is long compared to the
dynamical time scale of the component stars or planets. Once the
foundation of the potential is laid, we (i) solve for the positions of
the three co-linear and two triangular Lagrangian points, (ii) study the
stability of the points, (iii) provide a semi-analytical fit for the
general solution of the volume-equivalent Roche lobe radius that allows
the determination of when mass transfer is initiated, and (iv) discuss
the role of the potential geometry and structure in inducing the
possibility of systemic mass loss. Throughout the analysis we compare
our results to the well-studied case of synchronous binaries with
circular orbits and quantify the differences, which turn out to be
significant depending on the binary properties. 

Our study has been motivated by the need for characterizing the
gravitational potential for the general case of eccentric,
non-synchronous binaries when considering interactions (mass transfer or
mass and angular momentum loss) that occur in binaries before full
synchronization and circularization has been achieved.  The analysis
presented here can be applied to both stellar and planetary systems for
a wide range of problems. We plan to use the results to study the
ballistic motion of test particles in the potential, assess the
conditions under which mass transfer may be conservative or not, and
eventually apply the results in the development of a theoretical
framework that allows modeling of the orbital evolution of interacting,
eccentric, non-synchronous binaries.


\acknowledgments
We thank Chris Belczynski and Fred Rasio for useful discussions, as well
as an anonymous referee for the encouragement to complete our analytic
fits for the Roche lobe radius..  This work is supported by a NASA GSRP
Fellowship to JS, a Packard Foundation Fellowship in Science and
Engineering and a NSF CAREER grant (AST-0449558) to VK. 

\bibliography{MT}

\end{document}